\documentclass[aps,prl,unsortedaddress,superscriptaddress,showpacs,nofootinbib,twocolumn,10pt]{revtex4-2}
\usepackage{natbib}
\usepackage[colorlinks=true, linkcolor=Blue,citecolor=Blue,urlcolor=Blue]{hyperref}
\usepackage[utf8]{inputenc}
\usepackage{color}
\usepackage[dvipsnames,x11names]{xcolor}
\usepackage{amsfonts,amsmath,amssymb}
\usepackage{graphicx}
\usepackage{bm}
\usepackage{epstopdf}
\usepackage{mathtools}
\usepackage{lipsum}

\newlength{\colw}
\usepackage{newtxtext}
\usepackage{newtxmath}
\usepackage[T1]{fontenc}

\usepackage{orcidlink}

\usepackage[normalem]{ulem}

\newcommand{\olsi}[1]{\,\overline{\!{#1}}} % overline short italic

\newcommand{\ific}{\affiliation{Instituto de F\'isica Corpuscular (centro mixto CSIC-UV), \\
Institutos de Investigaci\'on de Paterna, Apartado 22085, 46071, Valencia, Spain}}

\newcommand{\itp}{\affiliation{Institute of Theoretical Physics, Chinese Academy of Sciences, Beijing 100190, China}}

\newcommand{\ucas}{\affiliation{School of Physical Sciences, University of Chinese Academy of Sciences, Beijing 100049, China}}

\newcommand{\scnt}{\affiliation{Southern Center for Nuclear-Science Theory (SCNT), Institute of Modern Physics,\\ Chinese Academy of Sciences, Huizhou 516000, China}}

\usepackage{mfirstuc} % to uppercase the name
\newcommand{\addReviewer}[2]{
  \expandafter\newcommand\csname #1\endcsname[1]{{\color{#2}##1}}
  \expandafter\newcommand\csname #1cor\endcsname[2]{{\color{#2}\sout{##1}{{}##2}}}
  \expandafter\newcommand\csname #1color\endcsname{#2}
}
\addReviewer{Miguel}{DodgerBlue1}
\newcommand{\EndMatter}[1][End Matter]{%
  \par
  \onecolumngrid          % iguala las columnas, NO salta de página
  \vspace{1.2\baselineskip}
  \begin{center}
    \rule{0.45\textwidth}{0.4pt}\\[0.9\baselineskip]
    {\bfseries\MakeUppercase{#1}}
  \end{center}
  \vspace{0.4\baselineskip}
  \twocolumngrid          % vuelve a la rejilla de dos columnas
}
\makeatletter 
    
\renewcommand\onecolumngrid{% <<<<<<
\do@columngrid{one}{\@ne}%
\def\set@footnotewidth{\onecolumngrid}% <<<<<<<<<<<<<<<<
\def\footnoterule{\kern-6pt\hrule width 1.5in\kern6pt}%
}

\renewcommand\twocolumngrid{% <<<<<<
        \def\footnoterule{% restore rule
        \dimen@\skip\footins\divide\dimen@\thr@@
        \kern-\dimen@\hrule width.5in\kern\dimen@}
        \do@columngrid{mlt}{\tw@}
}%

\makeatother    
\usepackage[compact]{titlesec}
\titleformat{\section}[runin]{\itshape}{\noindent\thesection.}{}{}[---]
\titlespacing*{\section}
  {0pt}   % sangría izquierda
  {*1}    % espacio antes
  {0pt}   % separación después (run-in)
\usepackage{xparse}

\ExplSyntaxOn
\NewDocumentCommand{\OldText}{O{} m}
 {
  \str_if_eq:nnTF {#1} {remove}
    {}
    {{\color{Gray}#2}}
 }
\ExplSyntaxOff
\begin{document}

\title{\boldmath A femtoscopic tale of two $C$-parities: the $Z_c(3900)$ and the isovector partner of the $X(3872)$}

\author{Pan-Pan Shi\orcidlink{0000-0003-2057-9884}} \email{Panpan.Shi@ific.uv.es}
\ific

\author{Miguel Albaladejo\orcidlink{0000-0001-7340-9235}}\email{Miguel.Albaladejo@ific.uv.es}
\ific 

\author{Feng-Kun~Guo\orcidlink{0000-0002-2919-2064}}\email{fkguo@itp.ac.cn}
\itp\ucas\scnt

\author{Juan Nieves\orcidlink{0000-0002-2518-4606}} \email{jmnieves@ific.uv.es}
\ific 

\frenchspacing

% \date{\today}

\begin{abstract}
Understanding the nature of the exotic $Z_c(3900)$ and $Z_{cs}(3985)$ states, and searching for the predicted isovector partner $W_{c1}$ of the $X(3872)$, remain central challenges in exotic-hadron spectroscopy. We investigate the femtoscopic correlation functions (CFs) of the $D^{(\ast)0}D_{(s)}^{(\ast)-}$ systems. Since these charm-meson--antimeson pairs are not $G$-parity eigenstates, their CFs contain contributions from both the $C$-odd and $C$-even sectors, providing direct access to the dynamics underlying the $Z_c$, $Z_{cs}$, and the predicted isovector exotic $W_{c1}$. Within a heavy-quark-spin-symmetric coupled-channel framework, we show that the $C$-even admixture enhances the low-momentum CFs by more than $2.5\sigma$ in the vicinity of their thresholds. Free from Coulomb distortions and accessible in high-multiplicity $pp$ collisions at the LHC, these channels offer the first direct femtoscopic probe of the isovector $C$-even sector and of the elusive $W_{c1}$ state.
\end{abstract}

\maketitle

\section{Introduction}
Since the $X(3872)$, also known as $\chi_{c1}(3872)$ in the Review of Particle Physics~\cite{ParticleDataGroup:2026}, was first observed by the Belle Collaboration~\cite{Belle:2003nnu}, numerous structures have been reported in the hidden-charm sector. Many of these resonances are referred to as exotic states, since they challenge the traditional constituent quark model, in which mesons and baryons are described as quark–antiquark pairs and three-quark systems, respectively. Over the past two decades, extensive experimental and theoretical efforts have been devoted to understanding the nature of these XYZ states~\cite{Hosaka:2016pey,Esposito:2016noz,Lebed:2016hpi,Ali:2017jda,Olsen:2017bmm,Guo:2017jvc,Albuquerque:2018jkn,Liu:2019zoy,Guo:2019twa,Brambilla:2019esw,Chen:2022asf,Liu:2024uxn,Chen:2024eaq,Wang:2025sic,Dai:2026fkg}.

Among these states, the charged $Z_c(3900)^{\pm}$, first observed by the BESIII and Belle Collaborations~\cite{BESIII:2013ris,Belle:2013yex}, provide compelling evidence for an exotic state, since their minimal quark configuration would be $c\bar{c}u\bar{d}$ (for $Z_c(3900)^{+}$), as inferred from the final states in which they are detected, namely $J/\psi \pi^{\pm}$ and $D\olsi{D}{}^{\ast}/D^{\ast}\olsi{D}$. The experimental activity carried out \cite{Xiao:2013iha,BESIII:2015cld,BESIII:2017bua,D0:2018wyb,BESIII:2020qkh} includes the discovery of their neutral partner $Z_c(3900)^0$, thus completing the isospin triplet, and the determination of its spin-parity, $J^{P}=1^+$. Their strange partners, $Z_{cs}(3985)$, have also been observed near the $D\olsi{D}{}_s^{\ast }/\olsi{D}D_s^{\ast }$ threshold~\cite{BESIII:2020qkh,BESIII:2022qzr}. The isospin triplet $Z_{c}(3900)$ and the two doublets $Z_{cs}(3985)$ would almost fill a flavor octet~\cite{Ji:2022uie}.

The $S$-wave interactions of the $D\olsi{D}{}^{\ast }$ and $D^{\ast }\olsi{D}$ systems can be decomposed into sectors of definite $C$-parity~\cite{Hidalgo-Duque:2012rqv,Guo:2013sya}. The interaction in the isovector $C$-odd sector generates the $Z_c(3900)$ and, by SU(3)-flavor symmetry, also governs the $D\olsi{D}{}_s^{\ast }$ and $D^{\ast }\olsi{D}_s$ channels, where the $Z_{cs}(3985)$ states emerge. The $C$-even sector is equally intriguing, as a state denoted $W_{c1}$, the predicted isovector partner of the $X(3872)$, has recently been proposed in Refs.~\cite{Zhang:2024fxy,Ji:2025hjw}. Although experimental evidence for this state is still lacking, its existence is supported by a recent lattice quantum chromodynamics (QCD) calculation~\cite{Sadl:2024dbd}.

The $Z_{c}(3900)$ and $Z_{cs}(3985)$ have been extensively studied within a variety of theoretical frameworks, including compact tetraquarks~\cite{Braaten:2013boa,Dias:2013xfa,Maiani:2014aja,Qiao:2013raa,Wang:2020iqt,Maiani:2021tri,Wua:2023ntn,Shi:2021jyr,Wan:2020oxt,Wang:2020rcx,Jin:2020yjn}, $D\olsi{D}{}^{\ast }$ and $D\olsi{D}{}_s^{\ast }$ hadronic molecules~\cite{Wang:2013cya,Wilbring:2013cha,Guo:2013sya,Dong:2013iqa,Zhang:2013aoa,Gong:2016jzb,Guo:2014iya,Aceti:2014uea,Albaladejo:2015lob,Albaladejo:2016jsg,Du:2020vwb,Wang:2020dgr,Du:2022jjv,Chen:2026fnz,Yan:2023bwt,Chen:2023def,Yang:2020nrt,Meng:2021rdg,Sun:2020hjw,Wang:2020htx,Xu:2020evn,Yan:2021tcp,Ortega:2021enc,Wu:2021ezz,Baru:2021ddn}, as well as kinematical interpretations based on threshold enhancements or cusp effects~\cite{Swanson:2014tra,Swanson:2015bsa,Chen:2013wca,Chen:2013coa,Ikeno:2020mra,Pilloni:2016obd}. Nevertheless, a pronounced threshold cusp generally requires a nearby singularity~\cite{Guo:2014iya,Dong:2020hxe}. Although the production of the $Z_c(3900)$ is affected by a triangle singularity~\cite{Wang:2013hga,Wang:2013cya,Pilloni:2016obd,Guo:2019twa,Chen:2023def,Chen:2026fnz}, this mechanism alone cannot account for the observed structure~\cite{Guo:2014iya,Albaladejo:2015lob,Gong:2016jzb,Du:2022jjv,Chen:2023def,Chen:2026fnz}. A similar conclusion has been reached for the $Z_{cs}(3985)$~\cite{Yang:2020nrt,Baru:2021ddn}. Disentangling genuine near-threshold dynamics from kinematical effects is therefore essential for a reliable determination of the low-energy scattering parameters and, ultimately, for elucidating the nature of the $Z_c(3900)$ and $Z_{cs}(3985)$ states.

Despite these efforts, the nature of the $Z_{c}(3900)$ and $Z_{cs}(3985)$ states, as well as the search for the predicted $W_{c1}$, remain open issues, motivating the development of new experimental probes. Femtoscopy provides access to low-energy hadron interactions through two-particle momentum correlations~\cite{Koonin:1977fh,Lednicky:1981su,Pratt:1986cc,Pratt:1990zq,Bauer:1992ffu,Lednicky:1997qr,Lisa:2005dd,Morita:2014kza,Ohnishi:2016elb,Morita:2016auo,Hatsuda:2017uxk,Mihaylov:2018rva,Haidenbauer:2018jvl,Kamiya:2019uiw,Haidenbauer:2020uew,Haidenbauer:2021zvr,Vidana:2023olz,Albaladejo:2023pzq,Kamiya:2022thy,Liu:2024nac,Sarti:2023wlg,Molina:2023oeu,Albaladejo:2023wmv,Ikeno:2023ojl,Encarnacion:2024jge,Albaladejo:2024lam,Kamiya:2024diw,Torres-Rincon:2023qll,Barbat:2025orm,Liu:2023uly,Albaladejo:2025lhn,Encarnacion:2025luc,Encarnacion:2026iur}. Its central observable, the femtoscopic correlation function (CF), is defined as the ratio of the measured two-particle momentum distribution to the product of the corresponding single-particle distributions in high-multiplicity collisions, and is directly sensitive to the scattering amplitude through final-state interactions (FSIs). A rapidly growing body of experimental data is now available; see, \textit{e.g.}, Refs.~\cite{ALICE:2011kmy,ALICE:2018ysd,ALICE:2015hvw,ALICE:2019gcn,ALICE:2021mfm}, and Ref.~\cite{Liu:2024uxn} for a recent review of the interplay between femtoscopy and exotic hadrons. In particular, the ALICE Collaboration has measured the CFs of the $p\olsi{D}$ and $\olsi{p}D$ systems~\cite{ALICE:2022enj}, as well as those of the $D\pi$ and $DK$ systems~\cite{ALICE:2024bhk}, opening a new avenue for investigating charm-hadron interactions.

In the hidden-charm sector, only two closely related femtoscopic studies have been reported to date~\cite{Kamiya:2022thy,Liu:2024nac}. Kamiya \textit{et al.}~\cite{Kamiya:2022thy} investigated the $D^0\olsi{D}{}^{\ast 0}$ and $D^+D^{\ast -}$ systems associated with the $X(3872)$ using spherical Gaussian potentials including the Coulomb interaction, finding a pronounced enhancement of the $D^+D^{\ast -}$ CF at low relative momenta. More recently, Liu \textit{et al.}~\cite{Liu:2024nac} studied the $D^0D^{\ast -}$ and $D^0D_s^{\ast -}$ systems assuming the $Z_c(3900)$ and $Z_{cs}(3985)$ to be bound, virtual, or resonant states, and showed that each scenario leaves a distinctive imprint on the corresponding CFs.

Excitingly, as we demonstrate in this work, femtoscopy also provides access to the recently predicted $W_{c1}$ state. In the $e^+e^-\to Y(4260)\to D\olsi D{}^{\ast }\pi$ reaction~\cite{BESIII:2015pqw}, $G$-parity conservation suppresses the $C$-even contribution to the $J/\psi\pi$ amplitude, so that only the $C$-odd interaction associated with the $Z_c$ states is probed. In contrast, the charged $D\olsi D{}^{\ast }$ pairs measured in femtoscopic analyses are not $G$-parity eigenstates. Their CFs therefore receive contributions from both the $C$-odd interaction, which couples to $J/\psi\pi$, and the $C$-even interaction in the heavy-meson sector. Consequently, $D\olsi D{}^{\ast }$ correlation measurements provide direct access to the $C$-even interaction and, in particular, to the predicted isovector partner $W_{c1}$ of the $X(3872)$~\cite{Zhang:2024fxy,Ji:2025hjw}. Furthermore, SU(3)-flavor symmetry relates the $C$-odd and $C$-even $D\olsi D{}^{\ast }$ interactions to those of the coupled $D^{\ast }\olsi D_s$--$D\olsi D{}_s^{\ast }$ system~\cite{Hidalgo-Duque:2012rqv}.\footnote{The $D^{\ast 0}D_s^-$ and $D^0D_s^{\ast -}$ channels are not related by $G$-parity but must be treated as coupled channels because their thresholds differ by only $1.8$~MeV. Using the masses from Ref.~\cite{ParticleDataGroup:2026}, the $D^0D^{\ast -}$ and $D^{\ast 0}D_s^{-}$ channels constitute the lowest thresholds in the $Z_c(3900)$ and $Z_{cs}(3985)$ sectors, respectively.}

In this Letter, we investigate the $D^{0}D^{\ast -}$, $D^{\ast 0}D^{-}$, $D^{0}D_s^{\ast -}$, and $D^{\ast 0}D_s^{-}$ CFs, demonstrating for the first time their sensitivity to the previously unexplored $C$-even interaction. Since these channels are free from Coulomb distortions~\cite{Haidenbauer:2021zvr,Torres-Rincon:2023qll,Albaladejo:2025kuv,Barbat:2026drc,Encarnacion:2026iur}, they provide particularly clean probes of the strong interaction.

\section{Formalism}\label{Sec:formala}

In the center-of-mass (CM) frame, the Koonin--Prat formula~\cite{Lisa:2005dd} for the CF observed in channel $i$ with on-shell momentum $k$, modified to account for coupled-channel effects \cite{Lednicky:1997qr,Haidenbauer:2018jvl,Vidana:2023olz}, is given by:
\begin{align}
{C}_{i}(k)=&1+4\pi\int_0^{+\infty}dr r^2\sum_jS_{j}(r)\nonumber\\
&\times\left(w_j|\tilde{\psi}_{ji}(k,r)|^2-j_0^2(kr)\delta_{ij}\right)\,,
\label{eq:CF}
\end{align}
where $w_j$ is the production weight of channel $j$ relative to the observation channel, normalized to $w_i=1$, and $j_0(kr)$ is the spherical Bessel function. The source $S_j(r)$ is usually parametrized by a Gaussian function as
\begin{align}
S_{j}(r)=\frac{1}{(\sqrt{4\pi})^{3}R_j^{3}}\text{exp}\left(-\frac{r^2}{4R_j^2}\right)\,.
\label{eq:source}
\end{align}
We use the effective radii $R_\text{eff}=0.84(7)$~fm for $pp$ collisions, and $2$ and $5$~fm for $pA$ and $AA$, respectively. Since $J/\psi$ production is strongly suppressed, we set $w_1=0$ for the $J/\psi\pi$ channel and assign unit weights to the $D^{(\ast)}\olsi{D}{}^{(\ast)}_{(s)}$ channels. Technical details are provided in \ref{Sec:Simulation}.

The $S$-wave relative wave function $\tilde\psi_{ji}(k,r)$ describing the transition from the intermediate channel $j$ to the observation channel $i$ is related to the scattering $T$-matrix by~\cite{Vidana:2023olz}
\begin{align}
\tilde\psi_{ji}(k,r) & = \delta_{ji} j_{0}(kr) + T_{ji}(E) \Theta(\Lambda-k) \nonumber\\
& \times \int_{0}^{\Lambda} \frac{q^2 dq}{2\pi^2} \frac{\omega_{1}^{j}+\omega_{2}^{j}}{2\omega_{1}^{j}\omega_{2}^{j}} \frac{j_0(qr)}{E^2-\left(\omega_{1}^{j}+\omega_{2}^{j}\right)^2+i\varepsilon},
\label{eq:wave_fuc}
\end{align}
where $E$ is the total CM energy of the hadron pair and $\omega_{\alpha}^{j}=(q^2+m_{\alpha,j}^2)^{1/2}$, with $m_{\alpha,j}$ the masses of the particles in the $j$-channel. We have assumed that the half off-shell $T-$matrix is expressed as
\begin{equation}
T_{ji}(E,p,q)= \Theta(\Lambda-p)T_{ji}(E)\Theta(\Lambda-q)\, ,
\label{eq:hos_scat}
\end{equation}
where $T_{ji}(E)$ is the corresponding on-shell scattering matrix element and $\Theta(\Lambda-k)$ denotes the  step function, with $\Lambda$ an ultraviolet (UV) cutoff used to renormalize the on-shell scattering amplitude~\cite{Gamermann:2009uq}.

As in the factorization employed for the $X(3872)$ production~\cite{Braaten:2005jj}, the CF separates into a short-distance contribution, encoded in the source $S_j(r)$, and a long-distance one described by the asymptotic wave function $\tilde\psi_{ji}(k,r)$. The Gaussian source in Eq.~\eqref{eq:source} acts as a regulator $\exp(-k^2R_j^2)$~\cite{Albaladejo:2024lam,Chen:2024eaq}. Two caveats follow. First, the cutoff $\Lambda$ in Eq.~\eqref{eq:wave_fuc} should, in principle, be absorbed into the source to ensure regulator-independent CFs~\cite{Chen:2024eaq,Epelbaum:2025aan,Molina:2025lzw}. Second, while Eq.~\eqref{eq:hos_scat} reproduces the on-shell $T$ matrix, it leaves the off-shell behavior undetermined, which is itself unobservable~\cite{Epelbaum:2025aan}. Thus, the short-distance source is interaction dependent and must be matched to the interaction rather than treated as universal. We retain the Gaussian source of Eq.~\eqref{eq:source} and estimate the associated uncertainty by varying $\Lambda$ between $0.6$ and $1.4$ GeV, which is expected to induce only small effects for realistic interactions~\cite{Molina:2025lzw}. By contrast, the scenario discrimination and the dependence on $C_{1X}$ discussed below are governed by on-shell amplitudes and are therefore much less sensitive to this systematic than the absolute CFs.

The coupled-channel scattering amplitudes for the $J/\psi \pi^-$--$D^0 D{}^{\ast -}$--$D^{\ast 0}D^{-}$ and $J/\psi K^-$--$D^{\ast 0} D_s^-$--$D^0D_s^{\ast -}$ systems in Eq.~\eqref{eq:wave_fuc} are derived from the Lippmann--Schwinger equation (LSE),
\begin{align}
T(E)=[1-V(E)G(E)]^{-1} \, V(E).
\label{eq:T_matrix}
\end{align}
Here $G(E)$ is a diagonal matrix whose elements are loop-functions $G_i(E)$, evaluated by a once-subtracted dispersion relation~\cite{Oller:1998zr}; see \textit{e.g.} Ref.\,\cite[Eq.\,(17)]{Du:2022jjv}. In particular, each loop function depends on a subtraction constant specified at a regularization scale $\mu$. We fix these constants $a_i(\mu=1~\text{GeV})$ to the values $-2.77$ for $i=1$, and $-2.5$ for $i=2,3$~\cite{Du:2022jjv}.\footnote{These subtraction constants, fixed in the state-of-the-art analyses of the $Z_c(3900)$ and $Z_{cs}(3985)$~\cite{Du:2022jjv} and of the $W_{c1}$~\cite{Ji:2025hjw}, are intimately related to the UV cutoff used in the wave function in Eq.~\eqref{eq:wave_fuc}~\cite{Nieves:2024dcz}. As discussed above, we conservatively estimate this systematic uncertainty by varying $\Lambda$.}

The effective potential matrix is constructed in terms of the heavy-quark spin (HQS) and light-flavor SU(3) symmetries~\cite{Hidalgo-Duque:2012rqv, Du:2022jjv,Du:2020vwb,Zhang:2024fxy,Ji:2025hjw},\footnote{The direct $J/\psi\pi$ and $J/\psi\olsi{K}$ transitions are Okubo-Zweig-Iizuka suppressed and are neglected in the present analysis. The smallness of the $J/\psi\pi$ scattering length was estimated in Ref.~\cite{Liu:2012dv}, and a recent analysis finds small scattering lengths for both $J/\psi\pi$ and $J/\psi K$~\cite{Yan:2026oil}.}
\begin{subequations}%
\label{eq:potential}
\begin{align}
V_{ij}(E) & = \mathcal{N}_{i} \widehat{V}_{ij}(E) \mathcal{N}_{j}\,,\quad \mathcal{N}_{i}=\sqrt{2m_{1,i}2m_{2,i}}\,,\\
\widehat{V}(E) & =\frac{1}{2}\begin{pmatrix}
0 & \sqrt{2} C_{12} & -\sqrt{2} C_{12}\\
\hphantom{+}\sqrt{2}C_{12} &C_{1X}+C^{\prime}_{1 Z} &  C_{1X}-C^{\prime}_{1Z} \\
-\sqrt{2}C_{12} &C_{1X}-C^{\prime}_{1Z} & C_{1X}+C^{\prime}_{1Z}
\end{pmatrix},
\end{align}
\end{subequations}
The low-energy constant (LEC) $C_{1X}$ parameterizes the contact interaction of the charmed-meson isovector pair in the $C$-even sector~\cite{Ji:2022vdj}. In the $C$-odd sector, we adopt the energy-dependent interaction extracted from the combined analysis of the $Z_c(3900)$ and $Z_{cs}(3985)$ exotic states performed in Ref.~\cite{Du:2022jjv},
\begin{align}\label{eq:V1Z-energy}
C_{1Z}^{\prime}=C_{1Z} + b\,(E^2-m_{\text{th}}^2)/(2m_{\text{th}})\,,
\end{align}
where $m_{\text{th}}$ denotes the average threshold of the last two channels ($D^{(\ast )}\olsi{D}{}_{(s)}^{(\ast )}$). 

Interestingly, as noted in the Introduction, the opposite signs of the $J/\psi\pi$ couplings to the two heavy-meson charge channels in Eq.~\eqref{eq:potential} imply that $J/\psi\pi$ couples only to the $C$-odd ($G$-even) combination $(D^0D^{\ast-}-D^{0}D^-)/\sqrt{2}$, while its coupling to the $C$-even ($G$-odd) combination $(D^0D^{-}+D^{\ast 0}D^-)/\sqrt{2}$ vanishes. Consequently, $C_{1X}$ influences the charged heavy-meson CFs only through the heavy-meson FSI, rather than via a direct $J/\psi\pi$ coupling. The resulting CFs in the physical basis therefore mix the $C$-odd ($C_{1Z}^{\prime}$) and $C$-even ($C_{1X}$) interaction components. This effect, neglected in previous studies~\cite{Liu:2024nac}, is predicted here for the first time.

\section{Fixing the theory parameters}

To quantify the impact of the $C$-even interaction on the $D^0D^{\ast -}$ and $D^{\ast 0}D_s^-$ CFs, we solve the full $J/\psi\pi^-$--$D^0D^{\ast -}$--$D^{\ast 0}D^-$ and $J/\psi K^-$--$D^{\ast 0}D_s^-$--$D^0D_s^{\ast -}$ coupled-channel systems, allowing for a nonzero $C_{12}$ in Eq.~\eqref{eq:potential}. The parameters $C_{12}$, $C_{1Z}$, and $b$ are taken from Ref.~\cite{Du:2022jjv}, where they were determined from a combined fit to the $J/\psi\pi^-$ and $D^0D^{\ast -}$ invariant-mass distributions in $e^+e^-\to J/\psi\pi^+\pi^-$~\cite{BESIII:2017bua} and $e^+e^-\to \pi^+D^0D^{\ast -}$~\cite{BESIII:2015pqw}, together with the $K^+$ recoil-mass distribution in $e^+e^-\to K^+(D^{\ast 0}D_s^-+D^0D_s^{\ast -})$~\cite{BESIII:2020qkh}. We use the two representative parameter sets (IIA and IIB) from Table~I of Ref.~\cite{Du:2022jjv}: $(C_{12},C_{1Z},b)=(0.006(1),-0.217(10),0)$ and $(0.005(1),-0.203(7),-0.473(45))$, in units of $(\mathrm{fm}^2,\mathrm{fm}^2,\mathrm{fm}^3)$, corresponding to the virtual-state and resonance interpretations of the $Z_c(3900)$ and $Z_{cs}(3985)$, respectively.\footnote{The recent analysis of Ref.~\cite{Chen:2026fnz}, which incorporates both experimental data and finite-volume lattice spectra~\cite{Cheung:2017tnt,CLQCD:2019npr,Sadl:2024dbd}, favors the resonance interpretation and supports the assignment of the $Z_c(3900)$ and $Z_{cs}(3985)$ to the same SU(3)-flavor octet~\cite{Yang:2020nrt}.} 

The $C$-even coupling $C_{1X}$ is fixed by reproducing the predicted pole position of the virtual state $W_{c1}^{\pm}$, the isovector partner of the $X(3872)$~\cite{Zhang:2024fxy,Ji:2025hjw}. The pole lies $8^{+8}_{-5}\,\mathrm{MeV}$ below the $D^0D^{\ast -}$ threshold, yielding $C_{1X}=-0.294(18)\,\mathrm{fm}^2$. Parameter uncertainties are propagated through bootstrap resampling, except for $C_{1X}$. Instead, its impact is examined separately in ~\ref{app:scattering-lengths}. Increasing the attraction moves the $W_{c1}$ pole toward threshold and strongly enhances the near-threshold CF, making it a sensitive probe of both the $C$-even interaction and the $W_{c1}$ state.

\section{Full-model prediction for the CFs}

\begin{figure}
\centering
\includegraphics[scale=0.75]{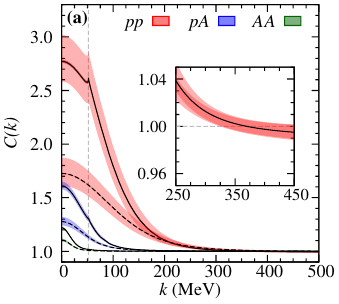}%
\includegraphics[scale=0.75]{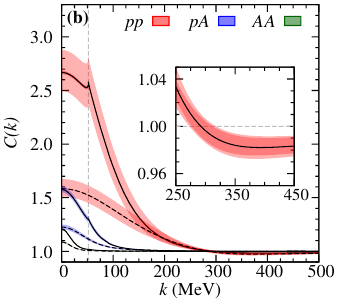}\\
\includegraphics[scale=0.75]{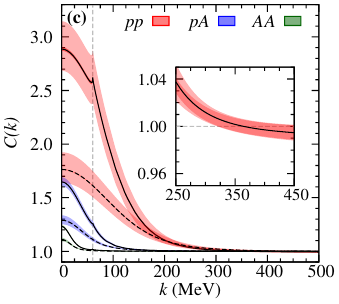}%
\includegraphics[scale=0.75]{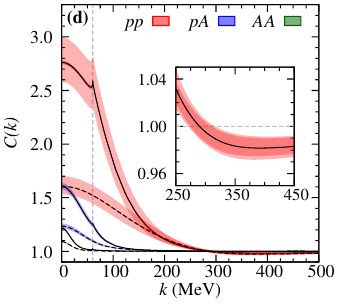}
\caption{CFs for the $D^0D^{\ast-}$ [(a),(b)] and $D^{\ast0}D_s^-$ [(c),(d)] channels. Vertical lines indicate the $D^{\ast0}D^-$ and $D^0D_s^{\ast-}$ thresholds. The left (right) panels correspond to the virtual-state (resonance) scenario for the $Z_c(3900)$ and $Z_{cs}(3985)$. Red, blue, and green denote $pp$, $pA$, and $AA$ collisions, respectively. Solid (dashed) lines show the results of the three-channel (two-channel) $T$-matrix calculations for the $J/\psi\pi^-$--$D^0D^{\ast-}$--$D^{\ast0}D^-$ and $J/\psi K^-$--$D^{\ast0}D_s^-$--$D^0D_s^{\ast-}$ ($J/\psi\pi^-$--$D^0D^{\ast-}$ and $J/\psi K^-$--$D^{\ast0}D_s^-$) systems. The outer bands represent the total uncertainty, obtained by combining the effects of the $1\sigma$ variations of the $pp$ source size, the LECs, and the UV cutoff $\Lambda$; the inner bands show the contribution from the cutoff variation alone.} 
\label{fig:CF_predict}
\end{figure}
\begin{figure}
\centering
\includegraphics[scale=0.75]{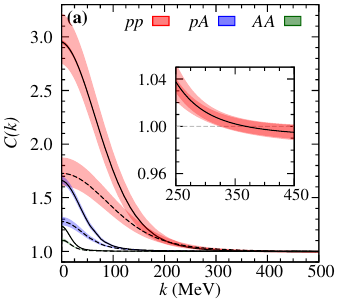}%
\includegraphics[scale=0.75]{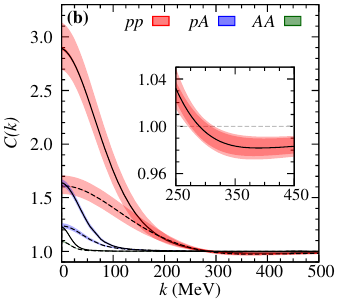}\\
\includegraphics[scale=0.75]{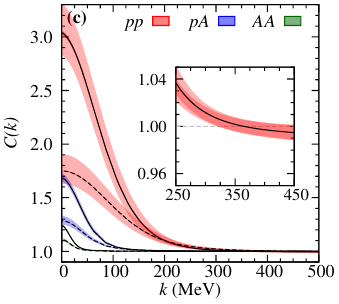}%
\includegraphics[scale=0.75]{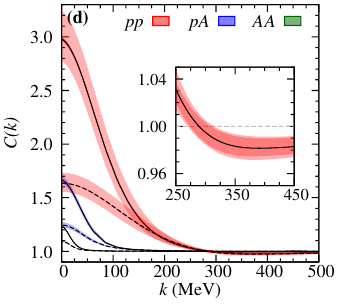}
\caption{Same as Fig.~\ref{fig:CF_predict}, but for the $D^{\ast 0}D^-$ and $D^0D_s^{\ast-}$ channels, whose thresholds lie approximately $1.5$ MeV above those of the $D^0D^{\ast-}$ and $D^{\ast0}D_s^-$ channels.} 
\label{fig:CF_chan3}
\end{figure}

Figures~\ref{fig:CF_predict} and \ref{fig:CF_chan3} show our predictions for the $D^0D^{\ast -}$ and $D^{\ast 0}D_s^-$ CFs, and for the $D^{\ast 0}D^-$ and $D^{0}D_s^{\ast -}$ CFs, respectively. As both exhibit essentially the same behavior, we discuss them jointly. The only qualitative difference is the mild threshold enhancement seen in Fig.~\ref{fig:CF_predict}, which is absent in Fig.~\ref{fig:CF_chan3}.

As shown in Fig.~\ref{fig:CF_predict}, the full three-channel $D^0D^{\ast-}$ and $D^{\ast0}D_s^-$ CFs (solid lines) are significantly enhanced above unity at threshold owing to the combined effect of the $Z_{c}$, $Z_{cs}$, $W_{c1}$, and $W_{cs1}$ states, and decrease with relative momentum $k$ in both the virtual-state and resonance scenarios. This threshold enhancement is reproduced to within a few percent by Eq.~(A5) of Ref.~\cite{Albaladejo:2023pzq}, which relates the CF at threshold to the scattering length within the Lednicky--Lyuboshits approximation~\cite{Lednicky:1981su}. Further discussion is given in \ref{app:scattering-lengths}.

In the virtual-state scenario (left panels), the central CFs remain mostly above unity over the entire momentum range for all source radii considered. By contrast, in the resonance scenario (right panels), they fall below unity for $k\gtrsim0.29$ GeV in $pp$ collisions. The corresponding minima occur at momenta slightly above the resonance pole positions, consistent with the Lednicky--Lyuboshits analysis of Ref.~\cite{Albaladejo:2023pzq}. Although the differences are small (see the zoomed insets), they are expected to be experimentally resolvable with the statistical precision projected for the ALICE 3 upgrade (see, \textit{e.g.}, Ref.~\cite[Fig.~46]{ALICE:2022wwr}). Together with the availability of four independent channels, this should allow the two scenarios to be distinguished. This constitutes one of the main predictions of our work.

To isolate the effect of the $C$-even interaction, we compare the full three-channel results with a two-channel calculation retaining a single heavy-meson channel, namely $J/\psi\pi^-$--$D^0D^{\ast-}$ and $J/\psi K^-$--$D^{\ast0}D_s^-$. The corresponding results are shown in Fig.~\ref{fig:CF_predict} as solid and dashed lines, respectively. This reference calculation is obtained from Eq.~\eqref{eq:potential} by setting $C_{1X}=C_{1Z}^{\prime}$, for which the off-diagonal potential $V_{23}\propto C_{1X}-C_{1Z}^{\prime}$ vanishes, the two heavy-meson channels decouple, and the diagonal interactions reduce to $C_{1Z}^{\prime}$. We emphasize that this condition does not project onto the $C$-odd sector. Rather, it makes the $C$-even and $C$-odd sectors degenerate; in contrast, a genuine $C$-odd projection would retain the combination $(D^0D^{\ast-}-D^{\ast0}D^-)/\sqrt{2}$ without requiring $C_{1X}=C_{1Z}^{\prime}$. Within this two-channel approximation we employ the effective potential $(C_{12},C_{1Z}^{\prime})$ of Ref.~\cite{Du:2022jjv}. This scheme has the same channel content as Ref.~\cite{Liu:2024nac}, but adopts the $C$-odd interaction extracted in Ref.~\cite{Du:2022jjv} from a combined fit to invariant-mass spectra, rather than determining it from the $Z_c(3900)$ and $Z_{cs}(3985)$ pole positions in the $C$-odd sector, as in Ref.~\cite{Liu:2024nac}.%

Our main result is that the three-channel CFs exceed their two-channel counterparts by more than $2.5\sigma$ near threshold in both the virtual-state and resonance scenarios. The $D^0D^{\ast-}$ and $D^{\ast0}D_s^-$ CFs therefore encode not only the $C$-odd dynamics associated with the $Z_c(3900)$ and $Z_{cs}(3985)$, but also sizable $C$-even contributions, neglected in Ref.~\cite{Liu:2024nac} and predicted here for the first time. Charmed-meson--antimeson femtoscopy thus provides direct access to the isovector $C$-even interaction, opening a new avenue for the experimental search for the predicted $W_{c1}$~\cite{Zhang:2024fxy,Ji:2025hjw}.

From Eq.~\eqref{eq:CF}, assuming degenerate thresholds and equal production weights, one finds
\begin{equation}\label{eq:relation-CFs-C-parity}
C_{D^0 D^{\ast-}}(k)=C_{D^{\ast0}D^-}(k)
=\big[\widetilde{C}_{1Z}(k)+\widetilde{C}_{1X}(k)\big]/2\,,
\end{equation}
where $\widetilde{C}_{1X,1Z}(k)$ denote the (unobservable) CFs of the (unphysical) $C$-even and $C$-odd eigenstates, respectively. Neglecting the $C$-even interaction yields $C_{D^0D^{\ast-}}(k)=\big[1+\widetilde{C}_{1Z}(k)\big]/2$. The results of Ref.~\cite{Liu:2024nac} may therefore be interpreted as the unphysical quantity $\widetilde{C}_{1Z}(k)$, rather than as the CF of the physical $D^0D^{\ast-}$ channel to which they are assigned. Under this interpretation, the observable CF would be $[1+\widetilde{C}_{1Z}(k)]/2$ instead of $\widetilde{C}_{1Z}(k)$, reducing by a factor of two the differences, and hence the discriminating power, among the resonance, virtual and bound state scenarios reported in Ref.~\cite{Liu:2024nac}. Moreover, in the physical case $\widetilde{C}_{1X}(k)\neq1$, the $C$-even interaction, which hosts the $W_{c1}$, contributes to the observable CF with the same weight as the $C$-odd interaction, as follows directly from Eq.~\eqref{eq:relation-CFs-C-parity}. Further details are provided in the Supplemental Material~\cite{supp}.

To assess the sensitivity to the cutoff in Eq.~\eqref{eq:wave_fuc}, we vary $\Lambda$ over the range $0.6$--$1.4$~GeV. The resulting CFs show only a mild cutoff dependence, reflected by the inner shaded bands in Fig.~\ref{fig:CF_predict}. Although this uncertainty becomes the dominant source at large momenta, its effect remains small: in $pp$ collisions, deviations from the central values never exceed $3\%$ in any scenario, and are even smaller for $pA$ and $AA$ collisions. These results are consistent with the estimates of Ref.~\cite{Molina:2025lzw} and demonstrate that, within the on-shell approach, the CFs are largely insensitive to the choice of cutoff.

\section{Summary and outlook}

We have shown that the femtoscopic $D^{(\ast)0}D_{(s)}^{(\ast)-}$ CFs simultaneously probe the $C$-odd and $C$-even sectors, which contribute to the measured CFs with equal weight. Using realistic source radii inferred from the average transverse mass, we have presented the first predictions including the $C$-even interaction, thereby substantially extending the analysis of Ref.~\cite{Liu:2024nac}, which considered only the $C$-odd sector associated with the $Z_c(3900)$ and $Z_{cs}(3985)$. The effect is sizable: the full three-channel CFs exceed their two-channel counterparts by more than $2.5\sigma$ at threshold in every system. In addition, the cusps generated by the opening of the coupled channels, located at $k\simeq50$--$60\,\mathrm{MeV}$ in the lowest-threshold CFs, provide distinctive signatures absent in single-channel descriptions, while the virtual-state and resonance scenarios remain experimentally distinguishable in up to four independent channels.

The mechanism identified here is not specific to the present system. Femtoscopic studies of $D^{0}\olsi{D}{}^{\ast0}$ pairs should incorporate the coupled-channel dynamics of $D^{0}\olsi{D}{}^{\ast0}$--$D^{\ast0}\olsi{D}{}^{0}$--$D^{+}D^{\ast-}$--$D^{\ast+}D^{-}$, where the isoscalar and isovector, $C$-even and $C$-odd interactions all enter the physical CFs. The $C$-odd sector was neglected in Ref.~\cite{Kamiya:2022thy}, where its inclusion was already recognized as an important open issue. Our results show that the analogous effect is substantial in the isovector system studied here. Similar coupled-channel effects are expected in bottom-meson and light-meson systems, establishing femtoscopy as a unique probe of both $C$-parity sectors across a broad class of hadronic systems.%

Taken together, these results provide a compelling case for measuring the CFs of hadron pairs containing neutral hadrons, such as $D^{0}D^{\ast-}$ and $D^{\ast0}D_s^-$, which are free from Coulomb distortions that otherwise obscure the low-momentum signal~\cite{Barbat:2026drc}. A single measurement of these CFs would simultaneously determine the nature of the $Z_c(3900)$ and $Z_{cs}(3985)$ poles and provide the first experimental access to the predicted $W_{c1}$, the yet-unobserved isovector partner of the $X(3872)$.

\bigskip

\begin{acknowledgments}
The numerical calculations were performed on a dedicated workstation of the ETHNP group at IFIC and at the Southern Nuclear Science Computing Center (SNSC). This work is part of the Grants PID2023-147458NB-C21 and CEX2023-001292-S funded by MICIU/AEI/10.13039/501100011033 and by ERDF/EU, as well as of the PROMETEO program Grant CIPROM/2023/59 funded by Generalitat Valenciana 10.13039/501100003359. %
M.\,A.\,acknwoledges the \guillemotleft{}Ramón y Cajal\guillemotright{} program Grant RYC2022-038524-I funded by MICIU/AEI/10.13039/501100011033 and by ESF+, and the \guillemotleft{}Atracción de Talento\guillemotright{} program Grant PIE 20245AT019 funded by CSIC 10.13039/501100003339. %
The work of F.K.G. is supported in part by the National Key R\&D Program of China under Grant No.~2023YFA1606703; by the National Natural Science Foundation of China (NSFC) under Grants No.~12125507, No.~12361141819, and No.~12447101; and by the Chinese Academy of Sciences (CAS) under Grant No.~YSBR-101.
\end{acknowledgments}

\bibliographystyle{apsrev4-2-modified}

\bibliography{Zc_ref.bib}

@article{Mihaylov:2018rva,
    author = "Mihaylov, D. L. and Mantovani Sarti, V. and Arnold, O. W. and Fabbietti, L. and Hohlweger, B. and Mathis, A. M.",
    title = {{A femtoscopic Correlation Analysis Tool using the Schr\"odinger equation (CATS)}},
    eprint = "1802.08481",
    archivePrefix = "arXiv",
    primaryClass = "hep-ph",
    doi = "10.1140/epjc/s10052-018-5859-0",
    journal = "Eur. Phys. J. C",
    volume = "78",
    number = "5",
    pages = "394",
    year = "2018"
}

@article{Morita:2014kza,
    author = "Morita, Kenji and Furumoto, Takenori and Ohnishi, Akira",
    title = "{$\Lambda\Lambda$ interaction from relativistic heavy-ion collisions}",
    eprint = "1408.6682",
    archivePrefix = "arXiv",
    primaryClass = "nucl-th",
    reportNumber = "YITP-14-67",
    doi = "10.1103/PhysRevC.91.024916",
    journal = "Phys. Rev. C",
    volume = "91",
    number = "2",
    pages = "024916",
    year = "2015"
}

@article{Ohnishi:2016elb,
    author = "Ohnishi, Akira and Morita, Kenji and Miyahara, Kenta and Hyodo, Tetsuo",
    title = "{Hadron\textendash{}hadron correlation and interaction from heavy\textendash{}ion collisions}",
    eprint = "1603.05761",
    archivePrefix = "arXiv",
    primaryClass = "nucl-th",
    reportNumber = "YITP-16-37, KUNS-2614",
    doi = "10.1016/j.nuclphysa.2016.05.010",
    journal = "Nucl. Phys. A",
    volume = "954",
    pages = "294--307",
    year = "2016"
}

@article{Morita:2016auo,
    author = "Morita, Kenji and Ohnishi, Akira and Etminan, Faisal and Hatsuda, Tetsuo",
    title = "{Probing multistrange dibaryons with proton-omega correlations in high-energy heavy ion collisions}",
    eprint = "1605.06765",
    archivePrefix = "arXiv",
    primaryClass = "hep-ph",
    reportNumber = "YITP-16-62, RIKEN-QHP-223",
    doi = "10.1103/PhysRevC.94.031901",
    journal = "Phys. Rev. C",
    volume = "94",
    number = "3",
    pages = "031901",
    year = "2016",
    note = "[Erratum: Phys.Rev.C 100, 069902 (2019)]"
}

@article{Hatsuda:2017uxk,
    author = "Hatsuda, Tetsuo and Morita, Kenji and Ohnishi, Akira and Sasaki, Kenji",
    editor = "Heinz, Ulrich and Evdokimov, Olga and Jacobs, Peter",
    title = "{$p\Xi^-$ Correlation in Relativistic Heavy Ion Collisions with Nucleon-Hyperon Interaction from Lattice QCD}",
    eprint = "1704.05225",
    archivePrefix = "arXiv",
    primaryClass = "nucl-th",
    reportNumber = "RIKEN-QHP-307",
    doi = "10.1016/j.nuclphysa.2017.04.041",
    journal = "Nucl. Phys. A",
    volume = "967",
    pages = "856--859",
    year = "2017"
}

@article{Kamiya:2019uiw,
    author = "Kamiya, Yuki and Hyodo, Tetsuo and Morita, Kenji and Ohnishi, Akira and Weise, Wolfram",
    title = "{$K^-p$ Correlation Function from High-Energy Nuclear Collisions and Chiral SU(3) Dynamics}",
    eprint = "1911.01041",
    archivePrefix = "arXiv",
    primaryClass = "nucl-th",
    doi = "10.1103/PhysRevLett.124.132501",
    journal = "Phys. Rev. Lett.",
    volume = "124",
    number = "13",
    pages = "132501",
    year = "2020"
}

@article{Lednicky:1981su,
    author = "Lednicky, R. and Lyuboshits, V. L.",
    title = "{Final State Interaction Effect on Pairing Correlations Between Particles with Small Relative Momenta}",
    reportNumber = "JINR-E2-81-453",
    journal = "Yad. Fiz.",
    volume = "35",
    pages = "1316--1330",
    year = "1981",
    note = "{Sov. J. Nucl. Phys. \textbf{35}, 770 (1982).}",
}

@article{Koonin:1977fh,
    author = "Koonin, S. E.",
    title = "{Proton Pictures of High-Energy Nuclear Collisions}",
    doi = "10.1016/0370-2693(77)90340-9",
    journal = "Phys. Lett. B",
    volume = "70",
    pages = "43--47",
    year = "1977"
}

@article{Pratt:1990zq,
    author = "Pratt, S. and Csorgo, T. and Zimanyi, J.",
    title = "{Detailed predictions for two pion correlations in ultrarelativistic heavy ion collisions}",
    doi = "10.1103/PhysRevC.42.2646",
    journal = "Phys. Rev. C",
    volume = "42",
    pages = "2646--2652",
    year = "1990"
}

@article{Bauer:1992ffu,
    author = "Bauer, W. and Gelbke, C. K. and Pratt, S.",
    title = "{Hadronic interferometry in heavy ion collisions}",
    doi = "10.1146/annurev.ns.42.120192.000453",
    journal = "Ann. Rev. Nucl. Part. Sci.",
    volume = "42",
    pages = "77--100",
    year = "1992"
}

@article{Pratt:1986cc,
    author = "Pratt, S.",
    title = "{Pion Interferometry of Quark-Gluon Plasma}",
    doi = "10.1103/PhysRevD.33.1314",
    journal = "Phys. Rev. D",
    volume = "33",
    pages = "1314--1327",
    year = "1986"
}

@article{Encarnacion:2025luc,
    author = "Encarnaci{\'o}n, P. and Albaladejo, M. and Feijoo, A. and Nieves, J.",
    title = "{Spectroscopic and femtoscopic insights into vector{\textendash}baryon interactions in the strangeness $-1$ sector}",
    eprint = "2507.08466",
    archivePrefix = "arXiv",
    primaryClass = "hep-ph",
    doi = "10.1140/epjc/s10052-025-14806-6",
    journal = "Eur. Phys. J. C",
    volume = "85",
    number = "11",
    pages = "1347",
    year = "2025"
}

@article{Sarti:2023wlg,
    author = "Sarti, V. Mantovani and Feijoo, A. and Vida{\~n}a, I. and Ramos, A. and Giacosa, F. and Hyodo, T. and Kamiya, Y.",
    title = "{Constraining the low-energy $S=-2$ meson-baryon interaction with two-particle correlations}",
    eprint = "2309.08756",
    archivePrefix = "arXiv",
    primaryClass = "hep-ph",
    doi = "10.1103/PhysRevD.110.L011505",
    journal = "Phys. Rev. D",
    volume = "110",
    number = "1",
    pages = "L011505",
    year = "2024"
}

@article{Encarnacion:2024jge,
    author = "Encarnaci{\'o}n, P. and Feijoo, A. and Sarti, V. Mantovani and Ramos, A.",
    title = "{Femtoscopic study of the $S=-1$ meson-baryon interaction: $K^{-}p$, $\pi^{-}\Lambda$, and $K^{+}\Xi^{-}$ correlations}",
    eprint = "2412.20880",
    archivePrefix = "arXiv",
    primaryClass = "hep-ph",
    doi = "10.1103/3ycr-vzmd",
    journal = "Phys. Rev. D",
    volume = "111",
    number = "11",
    pages = "114013",
    year = "2025",
    note = "[Erratum: Phys.Rev.D 113, 099901 (2026)]"
}

@article{Nieves:2024dcz,
    author = "Nieves, J. and Feijoo, A. and Albaladejo, M. and Du, Meng-Lin",
    title = "{Lowest-lying $\frac{1}{2}^{-}$ and $\frac{3}{2}^{-}$ $\Lambda_Q$ resonances: From the strange to the bottom sectors}",
    eprint = "2402.12726",
    archivePrefix = "arXiv",
    primaryClass = "hep-ph",
    doi = "10.1016/j.ppnp.2024.104118",
    journal = "Prog. Part. Nucl. Phys.",
    volume = "137",
    pages = "104118",
    year = "2024"
}

@article{Haidenbauer:2020uew,
    author = "Haidenbauer, J.",
    title = "{Exploring the $\Lambda$-deuteron interaction via correlations in heavy-ion collisions}",
    eprint = "2005.05012",
    archivePrefix = "arXiv",
    primaryClass = "nucl-th",
    doi = "10.1103/PhysRevC.102.034001",
    journal = "Phys. Rev. C",
    volume = "102",
    number = "3",
    pages = "034001",
    year = "2020"
}

@article{Haidenbauer:2021zvr,
    author = "Haidenbauer, Johann and Mei{\ss}ner, Ulf-G.",
    title = "{Exploring the $\Sigma^+p$ interaction by measurements of the correlation function}",
    eprint = "2109.11794",
    archivePrefix = "arXiv",
    primaryClass = "nucl-th",
    doi = "10.1016/j.physletb.2022.137074",
    journal = "Phys. Lett. B",
    volume = "829",
    pages = "137074",
    year = "2022"
}

@article{Kamiya:2024diw,
    author = "Kamiya, Yuki and Jinno, Asanosuke and Hyodo, Tetsuo and Ohnishi, Akira",
    title = "{Theoretical study of the $\Xi\alpha$ correlation function}",
    eprint = "2409.13207",
    archivePrefix = "arXiv",
    primaryClass = "nucl-th",
    reportNumber = "KUNS-3020",
    doi = "10.1103/284r-hvcg",
    journal = "Phys. Rev. C",
    volume = "113",
    number = "5",
    pages = "055205",
    year = "2026"
}

@article{Barbat:2025orm,
    author = "Barbat, Mikel F. and Torres-Rincon, Juan M. and Ramos, Angels and Tolos, Laura",
    title = "{Femtoscopy of DN and D{\textasciimacron}N systems}",
    eprint = "2507.07864",
    archivePrefix = "arXiv",
    primaryClass = "hep-ph",
    doi = "10.1103/6sqg-wtxd",
    journal = "Phys. Rev. D",
    volume = "113",
    number = "5",
    pages = "056025",
    year = "2026"
}

@article{Torres-Rincon:2023qll,
    author = "Torres-Rincon, Juan M. and Ramos, {\`A}ngels and Tolos, Laura",
    title = "{Femtoscopy of D mesons and light mesons upon unitarized effective field theories}",
    eprint = "2307.02102",
    archivePrefix = "arXiv",
    primaryClass = "hep-ph",
    doi = "10.1103/PhysRevD.108.096008",
    journal = "Phys. Rev. D",
    volume = "108",
    number = "9",
    pages = "096008",
    year = "2023"
}

@article{Albaladejo:2025kuv,
    author = "Albaladejo, M. and Garcia-Lorenzo, A. and Nieves, J.",
    title = "{Femtoscopy correlation functions and hadron-hadron scattering amplitudes in presence of Coulomb potential}",
    eprint = "2503.18710",
    archivePrefix = "arXiv",
    primaryClass = "hep-ph",
    doi = "10.1093/ptep/ptaf095",
    journal = "PTEP",
    volume = "2025",
    pages = "09",
    year = "2025"
}

@article{Barbat:2026drc,
    author = "Barbat, Mikel F. and Nieves, Juan and Tolos, Laura",
    title = "{Scattering and Femtoscopic Correlation Functions of the $\Sigma_c^{++}\pi^{+}$, $\Sigma_c^{0}\pi^{-}$ and $\Sigma_b^{+}\pi^{+}$ Systems}",
    eprint = "2603.02979",
    archivePrefix = "arXiv",
    primaryClass = "hep-ph",
    doi = "10.1016/j.physletb.2026.140529",
    journal = "Phys. Lett. B",
    volume = "878",
    pages = "140529",
    year = "2026"
}

@article{Liu:2023uly,
    author = "Liu, Zhi-Wei and Lu, Jun-Xu and Geng, Li-Sheng",
    title = "{Study of the DK interaction with femtoscopic correlation functions}",
    eprint = "2302.01046",
    archivePrefix = "arXiv",
    primaryClass = "hep-ph",
    doi = "10.1103/PhysRevD.107.074019",
    journal = "Phys. Rev. D",
    volume = "107",
    number = "7",
    pages = "074019",
    year = "2023"
}

@article{Chen:2026fnz,
    author = "Chen, Yun-Hua and Du, Meng-Lin and Guo, Feng-Kun",
    title = "{Determination of the $Z_c(3900)$ and the $Z_{cs}(3985)$ states from joint analysis of experimental and lattice data}",
    journal= "",
    eprint = "2604.25607",
    archivePrefix = "arXiv",
    primaryClass = "hep-ph",
    month = "4",
    year = "2026"
}

@article{Encarnacion:2026iur,
    author = "Encarnaci{\'o}n, Pablo and Garc{\'i}a-Lorenzo, Amador and Albaladejo, Miguel and Feijoo, Albert and Nieves, Juan and Vida{\~n}a, Isaac",
    title = "{Coulomb Effects in Momentum-Space Femtoscopy: A Case Study of the $\bar{K}\Omega$ System}",
    journal= "",
    eprint = "2607.11321",
    archivePrefix = "arXiv",
    primaryClass = "hep-ph",
    month = "7",
    year = "2026"
}

@article{Gamermann:2009uq,
    author = "Gamermann, D. and Nieves, J. and Oset, E. and Ruiz Arriola, E.",
    title = "{Couplings in coupled channels versus wave functions: application to the X(3872) resonance}",
    eprint = "0911.4407",
    archivePrefix = "arXiv",
    primaryClass = "hep-ph",
    doi = "10.1103/PhysRevD.81.014029",
    journal = "Phys. Rev. D",
    volume = "81",
    pages = "014029",
    year = "2010"
}

@article{Dai:2026fkg,
    author = "Dai, Xinchen and Jia, Sen and Nefediev, Alexey and Nieves, Juan and Shen, Chengping and Zhang, Liming",
    title = "{Exotic hadrons associated with $b$-quark}",
    journal ="",
    eprint = "2603.09315",
    archivePrefix = "arXiv",
    primaryClass = "hep-ph",
    month = "3",
    year = "2026"
}

@article{Hosaka:2016pey,
      author         = "Hosaka, Atsushi and Iijima, Toru and Miyabayashi,
                        Kenkichi and Sakai, Yoshihide and Yasui, Shigehiro",
      title          = "{Exotic hadrons with heavy flavors: $X, Y, Z$, and related
                        states}",
      journal        = "Prog. Theor. Exp. Phys.",
      volume         = "2016",
      year           = "2016",
      number         = "6",
      pages          = "062C01",
      doi            = "10.1093/ptep/ptw045",
      eprint         = "1603.09229",
      archivePrefix  = "arXiv",
      primaryClass   = "hep-ph",
      reportNumber   = "J-PARC-TH-0046",
      SLACcitation   = "%%CITATION = ARXIV:1603.09229;%%"
}

@article{Esposito:2016noz,
    author = "Esposito, A. and Pilloni, A. and Polosa, A. D.",
    title = "{Multiquark Resonances}",
    eprint = "1611.07920",
    archivePrefix = "arXiv",
    primaryClass = "hep-ph",
    reportNumber = "JLAB-THY-16-2301",
    doi = "10.1016/j.physrep.2016.11.002",
    journal = "Phys. Rept.",
    volume = "668",
    pages = "1--97",
    year = "2017"
}

@article{Lebed:2016hpi,
      author         = "Lebed, Richard F. and Mitchell, Ryan E. and Swanson, Eric
                        S.",
      title          = "{Heavy-quark QCD exotica}",
      journal        = "Prog. Part. Nucl. Phys.",
      volume         = "93",
      year           = "2017",
      pages          = "143-194",
      doi            = "10.1016/j.ppnp.2016.11.003",
      eprint         = "1610.04528",
      archivePrefix  = "arXiv",
      primaryClass   = "hep-ph",
      SLACcitation   = "%%CITATION = ARXIV:1610.04528;%%"
}

@article{Ali:2017jda,
      author         = "Ali, Ahmed and Lange, Jens S{\"o}ren and Stone, Sheldon",
      title          = "{Exotics: Heavy pentaquarks and tetraquarks}",
      journal        = "Prog. Part. Nucl. Phys.",
      volume         = "97",
      year           = "2017",
      pages          = "123-198",
      doi            = "10.1016/j.ppnp.2017.08.003",
      eprint         = "1706.00610",
      archivePrefix  = "arXiv",
      primaryClass   = "hep-ph",
      reportNumber   = "DESY-17-071",
      SLACcitation   = "%%CITATION = ARXIV:1706.00610;%%"
}

@article{Olsen:2017bmm,
      author         = "Olsen, Stephen Lars and Skwarnicki, Tomasz and Zieminska,
                        Daria",
      title          = "{Nonstandard heavy mesons and baryons: Experimental
                        evidence}",
      journal        = "Rev. Mod. Phys.",
      volume         = "90",
      year           = "2018",
      number         = "1",
      pages          = "015003",
      doi            = "10.1103/RevModPhys.90.015003",
      eprint         = "1708.04012",
      archivePrefix  = "arXiv",
      primaryClass   = "hep-ph",
      SLACcitation   = "%%CITATION = ARXIV:1708.04012;%%"
}

@article{Guo:2017jvc,
    author = "Guo, Feng-Kun and Hanhart, Christoph and Mei\ss{}ner, Ulf-G. and Wang, Qian and Zhao, Qiang and Zou, Bing-Song",
    title = "{Hadronic molecules}",
    eprint = "1705.00141",
    archivePrefix = "arXiv",
    primaryClass = "hep-ph",
    doi = "10.1103/RevModPhys.90.015004",
    journal = "Rev. Mod. Phys.",
    volume = "90",
    number = "1",
    pages = "015004",
    year = "2018",
    note = "[Erratum: Rev.Mod.Phys. 94, 029901 (2022)]"
}

@article{Albuquerque:2018jkn,
    author = "Albuquerque, Raphael M. and Dias, Jorgivan M. and Khemchandani, K. P. and Mart\'\i{}nez Torres, A. and Navarra, Fernando S. and Nielsen, Marina and Zanetti, Carina M.",
    title = "{QCD sum rules approach to the $X,~Y$ and $Z$ states}",
    eprint = "1812.08207",
    archivePrefix = "arXiv",
    primaryClass = "hep-ph",
    doi = "10.1088/1361-6471/ab2678",
    journal = "J. Phys. G",
    volume = "46",
    number = "9",
    pages = "093002",
    year = "2019"
}

@article{Liu:2019zoy,
    author = "Liu, Yan-Rui and Chen, Hua-Xing and Chen, Wei and Liu, Xiang and Zhu, Shi-Lin",
    title = "{Pentaquark and tetraquark states}",
    eprint = "1903.11976",
    archivePrefix = "arXiv",
    primaryClass = "hep-ph",
    doi = "10.1016/j.ppnp.2019.04.003",
    journal = "Prog. Part. Nucl. Phys.",
    volume = "107",
    pages = "237--320",
    year = "2019"
}

@article{Guo:2019twa,
    author = "Guo, Feng-Kun and Liu, Xiao-Hai and Sakai, Shuntaro",
    title = "{Threshold cusps and triangle singularities in hadronic reactions}",
    eprint = "1912.07030",
    archivePrefix = "arXiv",
    primaryClass = "hep-ph",
    doi = "10.1016/j.ppnp.2020.103757",
    journal = "Prog. Part. Nucl. Phys.",
    volume = "112",
    pages = "103757",
    year = "2020"
}

@article{Brambilla:2019esw,
    author = "Brambilla, Nora and Eidelman, Simon and Hanhart, Christoph and Nefediev, Alexey and Shen, Cheng-Ping and Thomas, Christopher E. and Vairo, Antonio and Yuan, Chang-Zheng",
    title = "{The $XYZ$ states: Experimental and theoretical status and perspectives}",
    eprint = "1907.07583",
    archivePrefix = "arXiv",
    primaryClass = "hep-ex",
    reportNumber = "TUM-EFT 125/19",
    doi = "10.1016/j.physrep.2020.05.001",
    journal = "Phys. Rep.",
    volume = "873",
    pages = "1--154",
    year = "2020"
}

@article{Chen:2022asf,
    author = "Chen, Hua-Xing and Chen, Wei and Liu, Xiang and Liu, Yan-Rui and Zhu, Shi-Lin",
    title = "{An updated review of the new hadron states}",
    eprint = "2204.02649",
    archivePrefix = "arXiv",
    primaryClass = "hep-ph",
    doi = "10.1088/1361-6633/aca3b6",
    journal = "Rept. Prog. Phys.",
    volume = "86",
    number = "2",
    pages = "026201",
    year = "2023"
}

@article{Liu:2024uxn,
    author = "Liu, Ming-Zhu and Pan, Ya-Wen and Liu, Zhi-Wei and Wu, Tian-Wei and Lu, Jun-Xu and Geng, Li-Sheng",
    title = "{Three ways to decipher the nature of exotic hadrons: Multiplets, three-body hadronic molecules, and correlation functions}",
    eprint = "2404.06399",
    archivePrefix = "arXiv",
    primaryClass = "hep-ph",
    doi = "10.1016/j.physrep.2024.12.001",
    journal = "Phys. Rept.",
    volume = "1108",
    pages = "1--108",
    year = "2025"
}

@article{Chen:2024eaq,
  title = {Production of exotic hadrons in {\emph{pp}} and nuclear collisions},
  author = {Chen, Jinhui and Guo, Feng-Kun and Ma, Yu-Gang and Shen, Cheng-Ping and Shou, Qiye and Wang, Qian and Wu, Jia-Jun and Zou, Bing-Song},
  year = {2025},
  month = feb,
  journal = {Nucl. Sci. Tech.},
  volume = {36},
  number = {4},
  eprint = {2411.18257},
  primaryclass = {hep-ph},
  pages = {55},
  doi = {10.1007/s41365-025-01664-w},
  urldate = {2024-11-28},
  archiveprefix = {arXiv}
}

@article{Wang:2025sic,
    author = "Wang, Zhi-Gang",
    title = "{Review of the QCD sum rules for exotic states}",
    eprint = "2502.11351",
    archivePrefix = "arXiv",
    primaryClass = "hep-ph",
    doi = "10.15302/frontphys.2026.016300",
    journal = "Front. Phys. (Beijing)",
    volume = "21",
    number = "1",
    pages = "016300",
    year = "2026"
}

@article{ParticleDataGroup:2026,
    author = "Takahashi, F. and others",
    collaboration = "Particle Data Group",
    title = "{Review of Particle Physics}",
    doi = "10.1142/S0217751X26300115",
    journal = "Int. J. Mod. Phys. A",
    volume = "41",
    pages = "2630011",
    year = "2026"
}

@article{Belle:2003nnu,
    author = "Choi, S. K. and others",
    collaboration = "Belle",
    title = "{Observation of a narrow charmonium-like state in exclusive $B^\pm \to K^\pm \pi^+ \pi^- J/\psi$ decays}",
    eprint = "hep-ex/0309032",
    archivePrefix = "arXiv",
    doi = "10.1103/PhysRevLett.91.262001",
    journal = "Phys. Rev. Lett.",
    volume = "91",
    pages = "262001",
    year = "2003"
}

@article{BESIII:2017bua,
    author = "Ablikim, Medina and others",
    collaboration = "BESIII",
    title = "{Determination of the Spin and Parity of the $Z_c(3900)$}",
    eprint = "1706.04100",
    archivePrefix = "arXiv",
    primaryClass = "hep-ex",
    doi = "10.1103/PhysRevLett.119.072001",
    journal = "Phys. Rev. Lett.",
    volume = "119",
    number = "7",
    pages = "072001",
    year = "2017"
}

@article{BESIII:2015pqw,
    author = "Ablikim, M. and others",
    collaboration = "BESIII",
    title = "{Confirmation of a charged charmoniumlike state $Z_c(3885)^{\mp}$ in $e^+e^-\to\pi^{\pm}(D\bar{D}^*)^\mp$ with double $D$ tag}",
    eprint = "1509.01398",
    archivePrefix = "arXiv",
    primaryClass = "hep-ex",
    doi = "10.1103/PhysRevD.92.092006",
    journal = "Phys. Rev. D",
    volume = "92",
    number = "9",
    pages = "092006",
    year = "2015"
}

@article{BESIII:2020qkh,
    author = "Ablikim, Medina and others",
    collaboration = "BESIII",
    title = "{Observation of a Near-Threshold Structure in the $K^+$ Recoil-Mass Spectra in $e^+e^- \rightarrow K^+(D_s^-D^{*0}+D_s^{*-}D^0$)}",
    eprint = "2011.07855",
    archivePrefix = "arXiv",
    primaryClass = "hep-ex",
    doi = "10.1103/PhysRevLett.126.102001",
    journal = "Phys. Rev. Lett.",
    volume = "126",
    number = "10",
    pages = "102001",
    year = "2021"
}

@article{BESIII:2013ris,
    author = "Ablikim, M. and others",
    collaboration = "BESIII",
    title = "{Observation of a Charged Charmoniumlike Structure in $e^+e^- \to \pi^+\pi^- J/\psi$ at $\sqrt{s}$ =4.26 GeV}",
    eprint = "1303.5949",
    archivePrefix = "arXiv",
    primaryClass = "hep-ex",
    doi = "10.1103/PhysRevLett.110.252001",
    journal = "Phys. Rev. Lett.",
    volume = "110",
    pages = "252001",
    year = "2013"
}

@article{Xiao:2013iha,
    author = "Xiao, T. and Dobbs, S. and Tomaradze, A. and Seth, Kamal K.",
    title = "{Observation of the Charged Hadron $Z_c^{\pm}(3900)$ and Evidence for the Neutral $Z_c^0(3900)$ in $e^+e^-\to \pi\pi J/\psi$ at $\sqrt{s}=4170$ MeV}",
    eprint = "1304.3036",
    archivePrefix = "arXiv",
    primaryClass = "hep-ex",
    doi = "10.1016/j.physletb.2013.10.041",
    journal = "Phys. Lett. B",
    volume = "727",
    pages = "366--370",
    year = "2013"
}

@article{D0:2018wyb,
    author = "Abazov, Victor Mukhamedovich and others",
    collaboration = "D0",
    title = "{Evidence for $Z_c^{\pm}(3900)$ in semi-inclusive decays of $b$-flavored hadrons}",
    eprint = "1807.00183",
    archivePrefix = "arXiv",
    primaryClass = "hep-ex",
    reportNumber = "FERMILAB-PUB-18-303-E",
    doi = "10.1103/PhysRevD.98.052010",
    journal = "Phys. Rev. D",
    volume = "98",
    number = "5",
    pages = "052010",
    year = "2018"
}

@article{BESIII:2015cld,
    author = "Ablikim, M. and others",
    collaboration = "BESIII",
    title = "{Observation of $Z_c(3900)^{0}$ in $e^+e^-\to\pi^0\pi^0 J/\psi$}",
    eprint = "1506.06018",
    archivePrefix = "arXiv",
    primaryClass = "hep-ex",
    doi = "10.1103/PhysRevLett.115.112003",
    journal = "Phys. Rev. Lett.",
    volume = "115",
    number = "11",
    pages = "112003",
    year = "2015"
}

@article{Belle:2013yex,
    author = "Liu, Z. Q. and others",
    collaboration = "Belle",
    title = "{Study of $e^+e^- \to \pi^+ \pi^- J/\psi$ and Observation of a Charged Charmoniumlike State at Belle}",
    eprint = "1304.0121",
    archivePrefix = "arXiv",
    primaryClass = "hep-ex",
    reportNumber = "BELLE-PREPRINT-2013-6, KEK-PREPRINT-2013-2",
    doi = "10.1103/PhysRevLett.110.252002",
    journal = "Phys. Rev. Lett.",
    volume = "110",
    pages = "252002",
    year = "2013",
    note = "[Erratum: Phys.Rev.Lett. 111, 019901 (2013)]"
}

@article{Cheung:2017tnt,
    author = "Cheung, Gavin K. C. and Thomas, Christopher E. and Dudek, Jozef J. and Edwards, Robert G.",
    collaboration = "Hadron Spectrum",
    title = "{Tetraquark operators in lattice QCD and exotic flavour states in the charm sector}",
    eprint = "1709.01417",
    archivePrefix = "arXiv",
    primaryClass = "hep-lat",
    reportNumber = "DAMTP-2017-33, JLAB-THY-17-2541",
    doi = "10.1007/JHEP11(2017)033",
    journal = "JHEP",
    volume = "11",
    pages = "033",
    year = "2017"
}

@article{CLQCD:2019npr,
    author = "Chen, Ting and Chen, Ying and Gong, Ming and Liu, Chuan and Liu, Liuming and Liu, Yu-Bin and Liu, Zhaofeng and Ma, Jian-Ping and Werner, Markus and Zhang, Jian-Bo",
    collaboration = "CLQCD",
    title = "{A coupled-channel lattice study on the resonance-like structure $Z_c(3900)$}",
    eprint = "1907.03371",
    archivePrefix = "arXiv",
    primaryClass = "hep-lat",
    doi = "10.1088/1674-1137/43/10/103103",
    journal = "Chin. Phys. C",
    volume = "43",
    number = "10",
    pages = "103103",
    year = "2019"
}

@article{Braaten:2013boa,
    author = "Braaten, Eric",
    title = "{How the $Z_c$(3900) Reveals the Spectra of Quarkonium Hybrid and Tetraquark Mesons}",
    eprint = "1305.6905",
    archivePrefix = "arXiv",
    primaryClass = "hep-ph",
    doi = "10.1103/PhysRevLett.111.162003",
    journal = "Phys. Rev. Lett.",
    volume = "111",
    pages = "162003",
    year = "2013"
}

@article{Dias:2013xfa,
    author = "Dias, J. M. and Navarra, F. S. and Nielsen, M. and Zanetti, C. M.",
    title = "{$Z^+_c$(3900) decay width in QCD sum rules}",
    eprint = "1304.6433",
    archivePrefix = "arXiv",
    primaryClass = "hep-ph",
    doi = "10.1103/PhysRevD.88.016004",
    journal = "Phys. Rev. D",
    volume = "88",
    number = "1",
    pages = "016004",
    year = "2013"
}

@article{Maiani:2014aja,
    author = "Maiani, L. and Piccinini, F. and Polosa, A. D. and Riquer, V.",
    title = "{The Z(4430) and a New Paradigm for Spin Interactions in Tetraquarks}",
    eprint = "1405.1551",
    archivePrefix = "arXiv",
    primaryClass = "hep-ph",
    doi = "10.1103/PhysRevD.89.114010",
    journal = "Phys. Rev. D",
    volume = "89",
    pages = "114010",
    year = "2014"
}

@article{Qiao:2013raa,
    author = "Qiao, Cong-Feng and Tang, Liang",
    title = "{Estimating the mass of the hidden charm $1^+(1^{+})$ tetraquark state via QCD sum rules}",
    eprint = "1307.6654",
    archivePrefix = "arXiv",
    primaryClass = "hep-ph",
    doi = "10.1140/epjc/s10052-014-3122-x",
    journal = "Eur. Phys. J. C",
    volume = "74",
    number = "10",
    pages = "3122",
    year = "2014"
}

@article{Wang:2020iqt,
    author = "Wang, Zhi-Gang",
    title = "{Analysis of $Z_{cs}(3985)$ as the axialvector tetraquark state}",
    eprint = "2011.10959",
    archivePrefix = "arXiv",
    primaryClass = "hep-ph",
    doi = "10.1088/1674-1137/abfa83",
    journal = "Chin. Phys. C",
    volume = "45",
    number = "7",
    pages = "073107",
    year = "2021"
}

@article{Wua:2023ntn,
    author = "Wua, Ren-Hua and Wang, Chen-Yu and Meng, Ce and Ma, Yan-Qing and Chao, Kuang-Ta",
    title = "{Z$_{c}$ and Z$_{cs}$ systems with operator mixing at NLO in QCD sum rules}",
    eprint = "2312.14224",
    archivePrefix = "arXiv",
    primaryClass = "hep-ph",
    doi = "10.1007/JHEP06(2024)216",
    journal = "JHEP",
    volume = "06",
    pages = "216",
    year = "2024"
}

@article{Shi:2021jyr,
    author = "Shi, Pan-Pan and Huang, Fei and Wang, Wen-Ling",
    title = "{Hidden charm tetraquark states in a diquark model}",
    eprint = "2105.02397",
    archivePrefix = "arXiv",
    primaryClass = "hep-ph",
    doi = "10.1103/PhysRevD.103.094038",
    journal = "Phys. Rev. D",
    volume = "103",
    number = "9",
    pages = "094038",
    year = "2021"
}

@article{Wan:2020oxt,
    author = "Wan, Bing-Dong and Qiao, Cong-Feng",
    title = "{About the exotic structure of $Z_{cs}$}",
    eprint = "2011.08747",
    archivePrefix = "arXiv",
    primaryClass = "hep-ph",
    doi = "10.1016/j.nuclphysb.2021.115450",
    journal = "Nucl. Phys. B",
    volume = "968",
    pages = "115450",
    year = "2021"
}

@article{Wang:2020rcx,
    author = "Wang, Qi-Nan and Chen, Wei and Chen, Hua-Xing",
    title = "{Exotic molecular states and tetraquark states with JP =0+, 1+, 2+}",
    eprint = "2011.10495",
    archivePrefix = "arXiv",
    primaryClass = "hep-ph",
    doi = "10.1088/1674-1137/ac0b3b",
    journal = "Chin. Phys. C",
    volume = "45",
    number = "9",
    pages = "093102",
    year = "2021"
}

@article{Jin:2020yjn,
    author = "Jin, Xin and Wu, Yuheng and Liu, Xuejie and Huang, Hongxia and Ping, Jialun and Zhong, Bin",
    title = "{Strange hidden-charm tetraquarks in constituent quark model}",
    eprint = "2011.12230",
    archivePrefix = "arXiv",
    primaryClass = "hep-ph",
    doi = "10.1140/epjc/s10052-021-09916-w",
    journal = "Eur. Phys. J. C",
    volume = "81",
    number = "12",
    pages = "1108",
    year = "2021"
}

@article{Wang:2013cya,
    author = "Wang, Qian and Hanhart, Christoph and Zhao, Qiang",
    title = "{Decoding the riddle of $Y(4260)$ and $Z_c(3900)$}",
    eprint = "1303.6355",
    archivePrefix = "arXiv",
    primaryClass = "hep-ph",
    doi = "10.1103/PhysRevLett.111.132003",
    journal = "Phys. Rev. Lett.",
    volume = "111",
    number = "13",
    pages = "132003",
    year = "2013"
}

@article{Wilbring:2013cha,
    author = "Wilbring, E. and Hammer, H. -W. and Mei\ss{}ner, U. -G.",
    title = "{Electromagnetic Structure of the $Z_c(3900)$}",
    eprint = "1304.2882",
    archivePrefix = "arXiv",
    primaryClass = "hep-ph",
    doi = "10.1016/j.physletb.2013.08.059",
    journal = "Phys. Lett. B",
    volume = "726",
    pages = "326--329",
    year = "2013"
}

@article{Guo:2013sya,
    author = "Guo, Feng-Kun and Hidalgo-Duque, Carlos and Nieves, Juan and Valderrama, Manuel Pavon",
    title = "{Consequences of Heavy Quark Symmetries for Hadronic Molecules}",
    eprint = "1303.6608",
    archivePrefix = "arXiv",
    primaryClass = "hep-ph",
    doi = "10.1103/PhysRevD.88.054007",
    journal = "Phys. Rev. D",
    volume = "88",
    pages = "054007",
    year = "2013"
}

@article{Dong:2013iqa,
    author = "Dong, Yubing and Faessler, Amand and Gutsche, Thomas and Lyubovitskij, Valery E.",
    title = "{Strong decays of molecular states Z$_{c}^{+}$ and Z$_{c}^{'+}$}",
    eprint = "1306.0824",
    archivePrefix = "arXiv",
    primaryClass = "hep-ph",
    doi = "10.1103/PhysRevD.88.014030",
    journal = "Phys. Rev. D",
    volume = "88",
    number = "1",
    pages = "014030",
    year = "2013"
}

@article{Zhang:2013aoa,
    author = "Zhang, Jian-Rong",
    title = "{Improved QCD sum rule study of $Z_{c}(3900)$ as a $\bar{D}D^{*}$ molecular state}",
    eprint = "1304.5748",
    archivePrefix = "arXiv",
    primaryClass = "hep-ph",
    doi = "10.1103/PhysRevD.87.116004",
    journal = "Phys. Rev. D",
    volume = "87",
    number = "11",
    pages = "116004",
    year = "2013"
}

@article{Aceti:2014uea,
    author = "Aceti, F. and Bayar, M. and Oset, E. and Martinez Torres, A. and Khemchandani, K. P. and Dias, Jorgivan Morais and Navarra, F. S. and Nielsen, M.",
    title = "{Prediction of an $I=1$ $D \bar D^*$ state and relationship to the claimed $Z_c(3900)$, $Z_c(3885)$}",
    eprint = "1401.8216",
    archivePrefix = "arXiv",
    primaryClass = "hep-ph",
    doi = "10.1103/PhysRevD.90.016003",
    journal = "Phys. Rev. D",
    volume = "90",
    number = "1",
    pages = "016003",
    year = "2014"
}

@article{Albaladejo:2016jsg,
    author = "Albaladejo, Miguel and Fernandez-Soler, Pedro and Nieves, Juan",
    title = "{$Z_c(3900)$: Confronting theory and lattice simulations}",
    eprint = "1606.03008",
    archivePrefix = "arXiv",
    primaryClass = "hep-ph",
    doi = "10.1140/epjc/s10052-016-4427-8",
    journal = "Eur. Phys. J. C",
    volume = "76",
    number = "10",
    pages = "573",
    year = "2016"
}

@article{Du:2020vwb,
    author = "Du, Meng-Chuan and Wang, Qian and Zhao, Qiang",
    title = "{The nature of charged charmonium-like states $Z_c(3900)$ and its strange partner $Z_{cs}(3982)$}",
    eprint = "2011.09225",
    archivePrefix = "arXiv",
    primaryClass = "hep-ph",
    journal = "",
    month = "11",
    year = "2020"
}

@article{Wang:2020dgr,
    author = "Wang, Zhi-Gang",
    title = "{Analysis of the Hidden-charm Tetraquark molecule mass spectrum with the QCD sum rules}",
    eprint = "2012.11869",
    archivePrefix = "arXiv",
    primaryClass = "hep-ph",
    doi = "10.1142/S0217751X21501074",
    journal = "Int. J. Mod. Phys. A",
    volume = "36",
    number = "15",
    pages = "2150107",
    year = "2021"
}

@article{Chen:2023def,
    author = "Chen, Yun-Hua and Du, Meng-Lin and Guo, Feng-Kun",
    title = "{Precise determination of the pole position of the exotic Z$_{c}(3900)$}",
    eprint = "2310.15965",
    archivePrefix = "arXiv",
    primaryClass = "hep-ph",
    doi = "10.1007/s11433-023-2408-1",
    journal = "Sci. China Phys. Mech. Astron.",
    volume = "67",
    number = "9",
    pages = "291011",
    year = "2024"
}

@article{Sadl:2024dbd,
    author = "Sadl, Mitja and Collins, Sara and Guo, Zhi-Hui and Padmanath, M. and Prelovsek, Sasa and Yan, Lin-Wan",
    title = "{Charmoniumlike channels $1^+$ with isospin 1 from lattice and effective field theory}",
    eprint = "2406.09842",
    archivePrefix = "arXiv",
    primaryClass = "hep-lat",
    doi = "10.1103/PhysRevD.111.054513",
    journal = "Phys. Rev. D",
    volume = "111",
    number = "5",
    pages = "054513",
    year = "2025"
}

@article{Guo:2014iya,
    author = "Guo, Feng-Kun and Hanhart, Christoph and Wang, Qian and Zhao, Qiang",
    title = "{Could the near-threshold $XYZ$ states be simply kinematic effects?}",
    eprint = "1411.5584",
    archivePrefix = "arXiv",
    primaryClass = "hep-ph",
    doi = "10.1103/PhysRevD.91.051504",
    journal = "Phys. Rev. D",
    volume = "91",
    number = "5",
    pages = "051504",
    year = "2015"
}

@article{Pilloni:2016obd,
    author = "Pilloni, A. and Fernandez-Ramirez, C. and Jackura, A. and Mathieu, V. and Mikhasenko, M. and Nys, J. and Szczepaniak, A. P.",
    collaboration = "JPAC",
    title = "{Amplitude analysis and the nature of the Z$_c$(3900)}",
    eprint = "1612.06490",
    archivePrefix = "arXiv",
    primaryClass = "hep-ph",
    reportNumber = "JLAB-THY-16-2410",
    doi = "10.1016/j.physletb.2017.06.030",
    journal = "Phys. Lett. B",
    volume = "772",
    pages = "200--209",
    year = "2017"
}

@article{Yang:2020nrt,
    author = "Yang, Zhi and Cao, Xu and Guo, Feng-Kun and Nieves, Juan and Valderrama, Manuel Pavon",
    title = "{Strange molecular partners of the $Z_c(3900)$ and $Z_c(4020)$}",
    eprint = "2011.08725",
    archivePrefix = "arXiv",
    primaryClass = "hep-ph",
    doi = "10.1103/PhysRevD.103.074029",
    journal = "Phys. Rev. D",
    volume = "103",
    number = "7",
    pages = "074029",
    year = "2021"
}

@article{Wang:2013hga,
    author = "Wang, Qian and Hanhart, Christoph and Zhao, Qiang",
    title = "{Systematic study of the singularity mechanism in heavy quarkonium decays}",
    eprint = "1305.1997",
    archivePrefix = "arXiv",
    primaryClass = "hep-ph",
    doi = "10.1016/j.physletb.2013.06.049",
    journal = "Phys. Lett. B",
    volume = "725",
    number = "1-3",
    pages = "106--110",
    year = "2013"
}

@article{Baru:2021ddn,
    author = "Baru, V. and Epelbaum, E. and Filin, A. A. and Hanhart, C. and Nefediev, A. V.",
    title = "{Is $Z_{cs}(3982)$ a molecular partner of $Z_c(3900)$ and $Z_c(4020)$ states?}",
    eprint = "2110.00398",
    archivePrefix = "arXiv",
    primaryClass = "hep-ph",
    doi = "10.1103/PhysRevD.105.034014",
    journal = "Phys. Rev. D",
    volume = "105",
    number = "3",
    pages = "034014",
    year = "2022"
}

@article{Sun:2020hjw,
    author = "Sun, Zhi-Feng and Xiao, Chu-Wen",
    title = "{Explanation of the newly obseaved $Z_{cs}^-(3985)$ as a $D_s^{(*)-}D^{(*)0}$ molecular state}",
    eprint = "2011.09404",
    archivePrefix = "arXiv",
    primaryClass = "hep-ph",
    journal = "",
    month = "11",
    year = "2020"
}

@article{Wang:2020htx,
    author = "Wang, Bo and Meng, Lu and Zhu, Shi-Lin",
    title = "{Decoding the nature of $Z_{cs}(3985)$ and establishing the spectrum of charged heavy quarkoniumlike states in chiral effective field theory}",
    eprint = "2011.10922",
    archivePrefix = "arXiv",
    primaryClass = "hep-ph",
    doi = "10.1103/PhysRevD.103.L021501",
    journal = "Phys. Rev. D",
    volume = "103",
    number = "2",
    pages = "L021501",
    year = "2021"
}

@article{Xu:2020evn,
    author = "Xu, Yong-Jiang and Liu, Yong-Lu and Cui, Chun-Yu and Huang, Ming-Qiu",
    title = "{$\bar D^{(*)}_s D^{(*)}$ molecular state with $J^P$= $1^+$}",
    eprint = "2011.14313",
    archivePrefix = "arXiv",
    primaryClass = "hep-ph",
    doi = "10.1103/PhysRevD.104.094028",
    journal = "Phys. Rev. D",
    volume = "104",
    number = "9",
    pages = "094028",
    year = "2021"
}

@article{Yan:2021tcp,
    author = "Yan, Mao-Jun and Peng, Fang-Zheng and S\'anchez S\'anchez, Mario and Pavon Valderrama, Manuel",
    title = "{Axial meson exchange and the $Z_c(3900)$ and $Z_{cs}(3985)$ resonances as heavy hadron molecules}",
    eprint = "2102.13058",
    archivePrefix = "arXiv",
    primaryClass = "hep-ph",
    doi = "10.1103/PhysRevD.104.114025",
    journal = "Phys. Rev. D",
    volume = "104",
    number = "11",
    pages = "114025",
    year = "2021"
}

@article{Ortega:2021enc,
    author = "Ortega, Pablo G. and Entem, David R. and Fernandez, F.",
    title = "{The strange partner of the $Z_c$ structures in a coupled-channels model}",
    eprint = "2103.07871",
    archivePrefix = "arXiv",
    primaryClass = "hep-ph",
    doi = "10.1016/j.physletb.2021.136382",
    journal = "Phys. Lett. B",
    volume = "818",
    pages = "136382",
    year = "2021"
}

@article{Gong:2016jzb,
    author = "Gong, Qin-Rong and Pang, Jing-Long and Wang, Yu-Fei and Zheng, Han-Qing",
    title = "{The $Z_c(3900)$ peak does not come from the \textquotedblleft{}triangle singularity\textquotedblright{}}",
    eprint = "1612.08159",
    archivePrefix = "arXiv",
    primaryClass = "hep-ph",
    doi = "10.1140/epjc/s10052-018-5690-7",
    journal = "Eur. Phys. J. C",
    volume = "78",
    number = "4",
    pages = "276",
    year = "2018"
}

@article{Wu:2021ezz,
    author = "Wu, Qi and Chen, Dian-Yong",
    title = "{Exploration of the hidden charm decays of Zcs(3985)}",
    eprint = "2108.06700",
    archivePrefix = "arXiv",
    primaryClass = "hep-ph",
    doi = "10.1103/PhysRevD.104.074011",
    journal = "Phys. Rev. D",
    volume = "104",
    number = "7",
    pages = "074011",
    year = "2021"
}

@article{Maiani:2021tri,
    author = "Maiani, Luciano and Polosa, Antonio D. and Riquer, Ver\'onica",
    title = "{The new resonances $Z_{cs}(3985)$ and $Z_{cs}(4003)$ (almost) fill two tetraquark nonets of broken $SU(3)_f$}",
    eprint = "2103.08331",
    archivePrefix = "arXiv",
    primaryClass = "hep-ph",
    doi = "10.1016/j.scib.2021.04.040",
    journal = "Sci. Bull.",
    volume = "66",
    pages = "1616--1619",
    year = "2021"
}

@article{Meng:2021rdg,
    author = "Meng, Lu and Wang, Bo and Wang, Guang-Juan and Zhu, Shi-Lin",
    title = "{Implications of the $Z_{cs}(3985)$ and $Z_{cs}(4000)$ as two different states}",
    eprint = "2104.08469",
    archivePrefix = "arXiv",
    primaryClass = "hep-ph",
    doi = "10.1016/j.scib.2021.06.026",
    journal = "Sci. Bull.",
    volume = "66",
    pages = "2065--2071",
    year = "2021"
}

@article{Du:2022jjv,
    author = "Du, Meng-Lin and Albaladejo, Miguel and Guo, Feng-Kun and Nieves, Juan",
    title = "{Combined analysis of the $Z_{c}(3900)$ and the $Z_{cs}(3985)$ exotic states}",
    eprint = "2201.08253",
    archivePrefix = "arXiv",
    primaryClass = "hep-ph",
    doi = "10.1103/PhysRevD.105.074018",
    journal = "Phys. Rev. D",
    volume = "105",
    number = "7",
    pages = "074018",
    year = "2022"
}

@article{Yan:2023bwt,
    author = "Yan, Lin-Wan and Guo, Zhi-Hui and Guo, Feng-Kun and Yao, De-Liang and Zhou, Zhi-Yong",
    title = "{Reconciling experimental and lattice data of $Z_c(3900)$ in a $J/\psi\,\pi^{-}$-$D\overline{D}{}^{\ast}$ coupled-channel analysis}",
    eprint = "2307.12283",
    archivePrefix = "arXiv",
    primaryClass = "hep-ph",
    doi = "10.1103/PhysRevD.109.014026",
    journal = "Phys. Rev. D",
    volume = "109",
    number = "1",
    pages = "014026",
    year = "2024"
}

@article{Swanson:2014tra,
    author = "Swanson, E. S.",
    title = "{$Z_b$ and $Z_c$ Exotic States as Coupled Channel Cusps}",
    eprint = "1409.3291",
    archivePrefix = "arXiv",
    primaryClass = "hep-ph",
    doi = "10.1103/PhysRevD.91.034009",
    journal = "Phys. Rev. D",
    volume = "91",
    number = "3",
    pages = "034009",
    year = "2015"
}

@article{Swanson:2015bsa,
    author = "Swanson, E. S.",
    title = "{Cusps and Exotic Charmonia}",
    eprint = "1504.07952",
    archivePrefix = "arXiv",
    primaryClass = "hep-ph",
    doi = "10.1142/S0218301316420106",
    journal = "Int. J. Mod. Phys. E",
    volume = "25",
    number = "07",
    pages = "1642010",
    year = "2016"
}

@article{Chen:2013wca,
    author = "Chen, Dian-Yong and Liu, Xiang and Matsuki, Takayuki",
    title = "{Predictions of Charged Charmoniumlike Structures with Hidden-Charm and Open-Strange Channels}",
    eprint = "1303.6842",
    archivePrefix = "arXiv",
    primaryClass = "hep-ph",
    doi = "10.1103/PhysRevLett.110.232001",
    journal = "Phys. Rev. Lett.",
    volume = "110",
    number = "23",
    pages = "232001",
    year = "2013"
}

@article{Chen:2013coa,
    author = "Chen, Dian-Yong and Liu, Xiang and Matsuki, Takayuki",
    title = "{Reproducing the $Z_c(3900)$ structure through the initial-single-pion-emission mechanism}",
    eprint = "1304.5845",
    archivePrefix = "arXiv",
    primaryClass = "hep-ph",
    doi = "10.1103/PhysRevD.88.036008",
    journal = "Phys. Rev. D",
    volume = "88",
    number = "3",
    pages = "036008",
    year = "2013"
}

@article{Ikeno:2020mra,
    author = "Ikeno, Natsumi and Molina, Raquel and Oset, Eulogio",
    title = "{The $Z_{cs}(3985)$ as a threshold effect from the $\bar D_s^* D + \bar D_s D^*$ interaction}",
    eprint = "2011.13425",
    archivePrefix = "arXiv",
    primaryClass = "hep-ph",
    doi = "10.1016/j.physletb.2021.136120",
    journal = "Phys. Lett. B",
    volume = "814",
    pages = "136120",
    year = "2021"
}

@article{ALICE:2024bhk,
    author = "Acharya, Shreyasi and others",
    collaboration = "ALICE",
    title = "{Studying the interaction between charm and light-flavor mesons}",
    eprint = "2401.13541",
    archivePrefix = "arXiv",
    primaryClass = "nucl-ex",
    reportNumber = "CERN-EP-2024-013",
    doi = "10.1103/PhysRevD.110.032004",
    journal = "Phys. Rev. D",
    volume = "110",
    number = "3",
    pages = "032004",
    year = "2024"
}

@article{ALICE:2020mfd,
    author = "Acharya, S. and others",
    collaboration = "ALICE",
    title = "{Unveiling the strong interaction among hadrons at the LHC}",
    eprint = "2005.11495",
    archivePrefix = "arXiv",
    primaryClass = "nucl-ex",
    reportNumber = "CERN-EP-2020-091",
    doi = "10.1038/s41586-020-3001-6",
    journal = "Nature",
    volume = "588",
    pages = "232--238",
    year = "2020",
    note = "[Erratum: Nature 590, E13 (2021)]"
}

@article{ALICE:2022enj,
    author = "Acharya, Shreyasi and others",
    collaboration = "ALICE",
    title = "{First study of the two-body scattering involving charm hadrons}",
    eprint = "2201.05352",
    archivePrefix = "arXiv",
    primaryClass = "nucl-ex",
    reportNumber = "CERN-EP-2022-006",
    doi = "10.1103/PhysRevD.106.052010",
    journal = "Phys. Rev. D",
    volume = "106",
    number = "5",
    pages = "052010",
    year = "2022"
}

@article{ALICE:2023sjd,
    author = "Acharya, Shreyasi and others",
    collaboration = "ALICE",
    title = "{Common femtoscopic hadron-emission source in pp collisions at the LHC}",
    eprint = "2311.14527",
    archivePrefix = "arXiv",
    primaryClass = "hep-ph",
    reportNumber = "CERN-EP-2023-267",
    doi = "10.1140/epjc/s10052-025-13793-y",
    journal = "Eur. Phys. J. C",
    volume = "85",
    number = "2",
    pages = "198",
    year = "2025",
    note = "[Erratum: Eur.Phys.J.C 86, 12 (2026)]"
}

@article{ALICE:2022wwr,
    author="{ALICE Collaboration}",
    primaryClass = "physics.ins-det",
    title = "{Letter of intent for ALICE 3: A next-generation heavy-ion experiment at the LHC}",
    eprint = "2211.02491",
    archivePrefix = "arXiv",
    reportNumber = "CERN-LHCC-2022-009, LHCC-I-038",
    journal = "",
    month = "11",
    year = "2022"
}

@article{Bierlich:2022pfr,
    author = "Bierlich, Christian and others",
    title = "{A comprehensive guide to the physics and usage of PYTHIA 8.3}",
    eprint = "2203.11601",
    archivePrefix = "arXiv",
    primaryClass = "hep-ph",
    reportNumber = "LU-TP 22-16, MCNET-22-04, FERMILAB-PUB-22-227-SCD",
    doi = "10.21468/SciPostPhysCodeb.8",
    journal = "SciPost Phys. Codeb.",
    volume = "2022",
    pages = "8",
    year = "2022"
}

@article{Skands:2014pea,
    author = "Skands, Peter and Carrazza, Stefano and Rojo, Juan",
    title = "{Tuning PYTHIA 8.1: the Monash 2013 Tune}",
    eprint = "1404.5630",
    archivePrefix = "arXiv",
    primaryClass = "hep-ph",
    reportNumber = "CERN-PH-TH-2014-069, MCNET-14-08, OUTP-14-05P",
    doi = "10.1140/epjc/s10052-014-3024-y",
    journal = "Eur. Phys. J. C",
    volume = "74",
    number = "8",
    pages = "3024",
    year = "2014"
}

@article{Lisa:2005dd,
    author = "Lisa, Michael Annan and Pratt, Scott and Soltz, Ron and Wiedemann, Urs",
    title = "{Femtoscopy in relativistic heavy ion collisions}",
    eprint = "nucl-ex/0505014",
    archivePrefix = "arXiv",
    doi = "10.1146/annurev.nucl.55.090704.151533",
    journal = "Ann. Rev. Nucl. Part. Sci.",
    volume = "55",
    pages = "357--402",
    year = "2005"
}

@article{Haidenbauer:2018jvl,
    author = "Haidenbauer, J.",
    title = "{Coupled-channel effects in hadron\textendash{}hadron correlation functions}",
    eprint = "1808.05049",
    archivePrefix = "arXiv",
    primaryClass = "hep-ph",
    doi = "10.1016/j.nuclphysa.2018.10.090",
    journal = "Nucl. Phys. A",
    volume = "981",
    pages = "1--16",
    year = "2019"
}

@article{Vidana:2023olz,
    author = "Vidana, I. and Feijoo, A. and Albaladejo, M. and Nieves, J. and Oset, E.",
    title = "{Femtoscopic correlation function for the $T_{cc}(3875)^+$ state}",
    eprint = "2303.06079",
    archivePrefix = "arXiv",
    primaryClass = "hep-ph",
    doi = "10.1016/j.physletb.2023.138201",
    journal = "Phys. Lett. B",
    volume = "846",
    pages = "138201",
    year = "2023"
}

@article{Albaladejo:2023pzq,
    author = "Albaladejo, Miguel and Nieves, Juan and Ruiz-Arriola, Enrique",
    title = "{Femtoscopic signatures of the lightest S-wave scalar open-charm mesons}",
    eprint = "2304.03107",
    archivePrefix = "arXiv",
    primaryClass = "hep-ph",
    doi = "10.1103/PhysRevD.108.014020",
    journal = "Phys. Rev. D",
    volume = "108",
    number = "1",
    pages = "014020",
    year = "2023"
}

@article{Kamiya:2022thy,
    author = "Kamiya, Yuki and Hyodo, Tetsuo and Ohnishi, Akira",
    title = "{Femtoscopic study on $DD^*$ and $D\bar{D}^*$ interactions for $T_{cc}$ and X(3872)}",
    eprint = "2203.13814",
    archivePrefix = "arXiv",
    primaryClass = "hep-ph",
    reportNumber = "YITP-22-26, RIKEN-iTHEMS-Report-22",
    doi = "10.1140/epja/s10050-022-00782-y",
    journal = "Eur. Phys. J. A",
    volume = "58",
    number = "7",
    pages = "131",
    year = "2022"
}

@article{Molina:2023oeu,
    author = "Molina, R. and Liu, Zhi-Wei and Geng, Li-Sheng and Oset, E.",
    title = "{Correlation function for the $a_0(980)$}",
    eprint = "2312.11993",
    archivePrefix = "arXiv",
    primaryClass = "hep-ph",
    doi = "10.1140/epjc/s10052-024-12694-w",
    journal = "Eur. Phys. J. C",
    volume = "84",
    number = "3",
    pages = "328",
    year = "2024"
}

@article{Albaladejo:2023wmv,
    author = "Albaladejo, M. and Feijoo, A. and Vida{\~n}a, I. and Nieves, J. and Oset, E.",
    title = "{Inverse problem in femtoscopic correlation functions: the $T_{cc}(3875)^+$ state}",
    eprint = "2307.09873",
    archivePrefix = "arXiv",
    primaryClass = "hep-ph",
    doi = "10.1140/epja/s10050-025-01650-1",
    journal = "Eur. Phys. J. A",
    volume = "61",
    number = "8",
    pages = "187",
    year = "2025"
}

@article{Ikeno:2023ojl,
    author = "Ikeno, Natsumi and Toledo, Genaro and Oset, Eulogio",
    title = "{Model independent analysis of femtoscopic correlation functions: An application to the $D^{\ast}_{s0}(2317)$}",
    eprint = "2305.16431",
    archivePrefix = "arXiv",
    primaryClass = "hep-ph",
    doi = "10.1016/j.physletb.2023.138281",
    journal = "Phys. Lett. B",
    volume = "847",
    pages = "138281",
    year = "2023"
}

@article{Albaladejo:2024lam,
    author = "Albaladejo, M. and Feijoo, A. and Nieves, J. and Oset, E. and Vida{\~n}a, I.",
    title = "{Femtoscopy correlation functions and mass distributions from production experiments}",
    eprint = "2410.08880",
    archivePrefix = "arXiv",
    primaryClass = "hep-ph",
    doi = "10.1103/PhysRevD.110.114052",
    journal = "Phys. Rev. D",
    volume = "110",
    number = "11",
    pages = "114052",
    year = "2024"
}

@article{Liu:2024nac,
    author = "Liu, Zhi-Wei and Lu, Jun-Xu and Liu, Ming-Zhu and Geng, Li-Sheng",
    title = "{Femtoscopy can tell whether Zc(3900) and Zcs(3985) are resonances, virtual states, or bound states}",
    eprint = "2404.18607",
    archivePrefix = "arXiv",
    primaryClass = "hep-ph",
    doi = "10.1016/j.scib.2025.09.022",
    journal = "Sci. Bull.",
    volume = "70",
    pages = "3515--3521",
    year = "2025"
}

@article{Albaladejo:2025lhn,
    author = "Albaladejo, Miguel and Canoa, Alejandro and Nieves, Juan and Pel{\'a}ez, Jose Ram{\'o}n and Ruiz-Arriola, Enrique and de Elvira, Jacobo Ruiz",
    title = "{The role of chiral symmetry and the non-ordinary $\kappa/K^{\ast}_{0}(700)$ nature in $\pi^{\pm}K_{S}$ femtoscopic correlations}",
    eprint = "2503.19746",
    archivePrefix = "arXiv",
    primaryClass = "hep-ph",
    reportNumber = "IPARCOS-UCM-25-020",
    doi = "10.1016/j.physletb.2025.139552",
    journal = "Phys. Lett. B",
    volume = "866",
    pages = "139552",
    year = "2025"
}

@article{Zhang:2024fxy,
    author = "Zhang, Zhen-Hua and Ji, Teng and Dong, Xiang-Kun and Guo, Feng-Kun and Hanhart, Christoph and Mei{\ss}ner, Ulf-G. and Rusetsky, Akaki",
    title = "{Predicting isovector charmonium-like states from X(3872) properties}",
    eprint = "2404.11215",
    archivePrefix = "arXiv",
    primaryClass = "hep-ph",
    doi = "10.1007/JHEP08(2024)130",
    journal = "JHEP",
    volume = "08",
    pages = "130",
    year = "2024"
}

@article{Ji:2025hjw,
    author = "Ji, Teng and Dong, Xiang-Kun and Guo, Feng-Kun and Hanhart, Christoph and Mei{\ss}ner, Ulf-G.",
    title = "{Precise determination of the properties of $X(3872)$ and of its isovector partner $W_{c1}$}",
    eprint = "2502.04458",
    archivePrefix = "arXiv",
    primaryClass = "hep-ph",
    journal = "",
    month = "2",
    year = "2025"
}

@article{Hidalgo-Duque:2012rqv,
    author = "Hidalgo-Duque, C. and Nieves, J. and Valderrama, M. Pavon",
    title = "{Light flavor and heavy quark spin symmetry in heavy meson molecules}",
    eprint = "1210.5431",
    archivePrefix = "arXiv",
    primaryClass = "hep-ph",
    doi = "10.1103/PhysRevD.87.076006",
    journal = "Phys. Rev. D",
    volume = "87",
    number = "7",
    pages = "076006",
    year = "2013"
}

@article{Braaten:2005jj,
    author = "Braaten, Eric and Kusunoki, Masaoki",
    title = "{Factorization in the production and decay of the X(3872)}",
    eprint = "hep-ph/0506087",
    archivePrefix = "arXiv",
    doi = "10.1103/PhysRevD.72.014012",
    journal = "Phys. Rev. D",
    volume = "72",
    pages = "014012",
    year = "2005"
}

@article{Ji:2022vdj,
    author = "Ji, Teng and Dong, Xiang-Kun and Albaladejo, Miguel and Du, Meng-Lin and Guo, Feng-Kun and Nieves, Juan and Zou, Bing-Song",
    title = "{Understanding the $0^{++}$ and $2^{++}$ charmonium(-like) states near 3.9~GeV}",
    eprint = "2212.00631",
    archivePrefix = "arXiv",
    primaryClass = "hep-ph",
    doi = "10.1016/j.scib.2023.02.034",
    journal = "Sci. Bull.",
    volume = "68",
    pages = "688--697",
    year = "2023"
}

@article{Albaladejo:2015lob,
    author = "Albaladejo, Miguel and Guo, Feng-Kun and Hidalgo-Duque, Carlos and Nieves, Juan",
    title = "{$Z_c(3900)$: What has been really seen?}",
    eprint = "1512.03638",
    archivePrefix = "arXiv",
    primaryClass = "hep-ph",
    doi = "10.1016/j.physletb.2016.02.025",
    journal = "Phys. Lett. B",
    volume = "755",
    pages = "337--342",
    year = "2016"
}

@article{Epelbaum:2025aan,
    author = "Epelbaum, Evgeny and Heihoff, Sven and Mei{\ss}ner, Ulf-G. and Tscherwon, Alexander",
    title = "{Can the Strong Interactions between Hadrons Be Determined Using Femtoscopy?}",
    eprint = "2504.08631",
    archivePrefix = "arXiv",
    primaryClass = "nucl-th",
    doi = "10.1103/tfsb-wlsd",
    journal = "Phys. Rev. Lett.",
    volume = "136",
    number = "21",
    pages = "212301",
    year = "2026"
}

@article{Yan:2026oil,
    author = "Yan, Jiang and Cao, Xiong-Hui and Du, Meng-Lin and Guo, Feng-Kun",
    title = "{Scattering lengths of the $J/\psi\pi$ and $J/\psi K$ systems}",
    eprint = "2601.18103",
    archivePrefix = "arXiv",
    primaryClass = "hep-ph",
    journal = "",
    month = "1",
    year = "2026"
}

@article{Oller:1998zr,
    author = "Oller, J. A. and Oset, E.",
    title = "{N/D description of two meson amplitudes and chiral symmetry}",
    eprint = "hep-ph/9809337",
    archivePrefix = "arXiv",
    doi = "10.1103/PhysRevD.60.074023",
    journal = "Phys. Rev. D",
    volume = "60",
    pages = "074023",
    year = "1999"
}

@article{CMS:2017mdg,
    author = "Sirunyan, Albert M and others",
    collaboration = "CMS",
    title = "{Bose-Einstein correlations in $pp$, $p\text{Pb}$, and $\text{Pb}\text{Pb}$ collisions at $\sqrt{{s}_{NN}}=0.9-7$ TeV}",
    eprint = "1712.07198",
    archivePrefix = "arXiv",
    primaryClass = "hep-ex",
    reportNumber = "CMS-FSQ-14-002, CERN-EP-2017-327",
    doi = "10.1103/PhysRevC.97.064912",
    journal = "Phys. Rev. C",
    volume = "97",
    number = "6",
    pages = "064912",
    year = "2018"
}

@article{ATLAS:2015dqi,
    author = "Aad, Georges and others",
    collaboration = "ATLAS",
    title = "{Two-particle Bose{\textendash}Einstein correlations in pp collisions at $\mathbf {\sqrt{s} =}$ 0.9 and 7 TeV measured with the ATLAS detector}",
    eprint = "1502.07947",
    archivePrefix = "arXiv",
    primaryClass = "hep-ex",
    reportNumber = "CERN-PH-EP-2014-264",
    doi = "10.1140/epjc/s10052-015-3644-x",
    journal = "Eur. Phys. J. C",
    volume = "75",
    number = "10",
    pages = "466",
    year = "2015"
}

@article{ALICE:2012yyu,
    author = "Abelev, Betty and others",
    collaboration = "ALICE",
    title = "{$K^0_s-K^0_s$ correlations in $pp$ collisions at $\sqrt{s}=7$ TeV from the LHC ALICE experiment}",
    eprint = "1206.2056",
    archivePrefix = "arXiv",
    primaryClass = "hep-ex",
    reportNumber = "CERN-PH-EP-2012-160",
    doi = "10.1016/j.physletb.2012.09.013",
    journal = "Phys. Lett. B",
    volume = "717",
    pages = "151--161",
    year = "2012"
}

@article{ALICE:2012aai,
    author = "Abelev, B. and others",
    collaboration = "ALICE",
    title = "{Charged kaon femtoscopic correlations in $pp$ collisions at $\sqrt{s}=7$ TeV}",
    eprint = "1212.5958",
    archivePrefix = "arXiv",
    primaryClass = "hep-ex",
    reportNumber = "CERN-PH-EP-2012-337",
    doi = "10.1103/PhysRevD.87.052016",
    journal = "Phys. Rev. D",
    volume = "87",
    number = "5",
    pages = "052016",
    year = "2013"
}

@article{CMS:2019fur,
    author = "Sirunyan, Albert M and others",
    collaboration = "CMS",
    title = "{Bose-Einstein correlations of charged hadrons in proton-proton collisions at $\sqrt{s} =$ 13 TeV}",
    eprint = "1910.08815",
    archivePrefix = "arXiv",
    primaryClass = "hep-ex",
    reportNumber = "CMS-FSQ-15-009, CERN-EP-2019-151",
    doi = "10.1007/JHEP03(2020)014",
    journal = "JHEP",
    volume = "03",
    pages = "014",
    year = "2020"
}

@article{ALICE:2020ibs,
    author = "Acharya, Shreyasi and others",
    collaboration = "ALICE",
    title = "{Search for a common baryon source in high-multiplicity pp collisions at the LHC}",
    eprint = "2004.08018",
    archivePrefix = "arXiv",
    primaryClass = "nucl-ex",
    reportNumber = "CERN-EP-2020-053",
    doi = "10.1016/j.physletb.2020.135849",
    journal = "Phys. Lett. B",
    volume = "811",
    pages = "135849",
    year = "2020",
    note = "[Erratum: Phys.Lett.B 861, 139233 (2025)]"
}

@article{Zhang:2024qkg,
    author = "Zhang, Zhen-Hua and Guo, Feng-Kun",
    title = "{Classification of coupled-channel near-threshold structures}",
    eprint = "2407.10620",
    archivePrefix = "arXiv",
    primaryClass = "hep-ph",
    doi = "10.1016/j.physletb.2025.139387",
    journal = "Phys. Lett. B",
    volume = "863",
    pages = "139387",
    year = "2025"
}

@article{Dong:2020hxe,
    author = "Dong, Xiang-Kun and Guo, Feng-Kun and Zou, Bing-Song",
    title = "{Explaining the Many Threshold Structures in the Heavy-Quark Hadron Spectrum}",
    eprint = "2011.14517",
    archivePrefix = "arXiv",
    primaryClass = "hep-ph",
    doi = "10.1103/PhysRevLett.126.152001",
    journal = "Phys. Rev. Lett.",
    volume = "126",
    number = "15",
    pages = "152001",
    year = "2021"
}

@article{Liu:2012dv,
    author = "Liu, Xiao-Hai and Guo, Feng-Kun and Epelbaum, Evgeny",
    title = "{Extracting $\pi\pi$ $S$-wave scattering lengths from cusp effect in heavy quarkonium dipion transitions}",
    eprint = "1212.4066",
    archivePrefix = "arXiv",
    primaryClass = "hep-ph",
    doi = "10.1140/epjc/s10052-013-2284-2",
    journal = "Eur. Phys. J. C",
    volume = "73",
    number = "1",
    pages = "2284",
    year = "2013"
}

@article{Molina:2025lzw,
    author = "Molina, R. and Oset, E.",
    title = "{Determination of off-shell ambiguities in correlation functions: Strategies to minimize them}",
    eprint = "2506.03669",
    archivePrefix = "arXiv",
    primaryClass = "hep-ph",
    doi = "10.1103/rst4-rkmm",
    journal = "Phys. Rev. D",
    volume = "112",
    number = "9",
    pages = "096006",
    year = "2025"
}

@article{ALICE:2023sgl,
    author = "Acharya, Shreyasi and others",
    collaboration = "ALICE",
    title = "{Charm production and fragmentation fractions at midrapidity in $pp$ collisions at $\sqrt{s}=13$ TeV}",
    eprint = "2308.04877",
    archivePrefix = "arXiv",
    primaryClass = "hep-ex",
    reportNumber = "CERN-EP-2023-162",
    doi = "10.1007/JHEP12(2023)086",
    journal = "JHEP",
    volume = "12",
    number = "12",
    pages = "086",
    year = "2023"
}

\EndMatter

\appendix

\newcounter{appsec}
\renewcommand{\theappsec}{Appendix \Alph{appsec}}
\newcommand{\appsection}[1]{\refstepcounter{appsec}\section{\theappsec.~#1}}

\appsection{Source radii and production weights}\label{Sec:Simulation}

The source radii $R_j$ are a key non-interaction input. It is approximately $1$~fm for $pp$ collisions and $5$ fm for heavy-ion collisions. In high-multiplicity (HM) $pp$ collisions the source radius follows a universal scaling with the pair average transverse mass $\langle m_T \rangle$~\cite{CMS:2017mdg,ATLAS:2015dqi,ALICE:2012yyu,ALICE:2012aai,CMS:2019fur,ALICE:2020ibs,ALICE:2023sjd}. Assuming the charmed-hadron pairs obey the same scaling, we evaluate $\langle m_T \rangle$ for each pair with \textsc{Pythia 8} at $\sqrt{s}=13$~TeV~\cite{Bierlich:2022pfr}, as detailed below.

The CFs are reconstructed from the HM events selected by the V0 detector in the ALICE Collaboration. The triggered events correspond to the first $0.17\%$ of inelastic $pp$ events in which at least one charged track is detected with pseudo-rapidity $|\eta|<1$. On average, about 30 charged particles are produced per collision event in the range $|\eta|<0.5$ in the HM environment~\cite{ALICE:2020mfd}. We neglect the charged particles produced from the FSI and the decays of unstable particles, and select the events that generate at least 15 charged particles within $|\eta|<0.5$ to emulate the HM environment. In experiments, the V0 detector records both the primary charged particles and those originating from decays of unstable particles. Therefore, requiring at least 15 charged particles provides a reasonable criterion for selecting hadron pairs of interest in an HM environment.

We simulate $pp$ collisions at a CM energy of $\sqrt s = 13$~TeV, including all hard QCD $2\to 2$ processes. Following Ref.\,\cite{ALICE:2022enj}, the Monash 2013 tune of \textsc{Pythia 8}~\cite{Skands:2014pea} is employed to set the QCD parameters. Charged pion and kaon candidates are selected with the transverse momentum ($p_{T}$) in the range of $[0.14,4.0]$ and $[0.15,2.15]$ GeV, respectively~\cite{ALICE:2024bhk}. No $p_T$ requirement is imposed on charmed mesons or charmonia in order to increase their statistics. These settings are chosen to reproduce the HM environment and to match the hadron-pair selection used by the ALICE Collaboration.

\begin{figure}[t]
\centering
\includegraphics[scale=0.75]{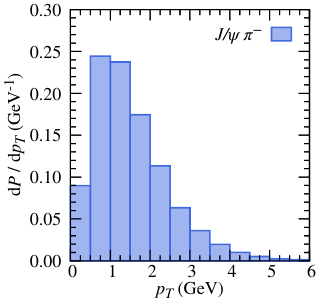}
\includegraphics[scale=0.75]{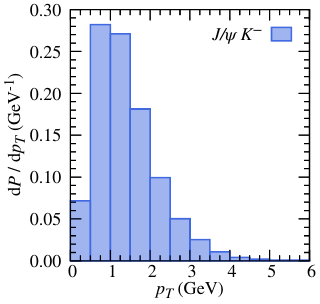}\\
\includegraphics[scale=0.75]{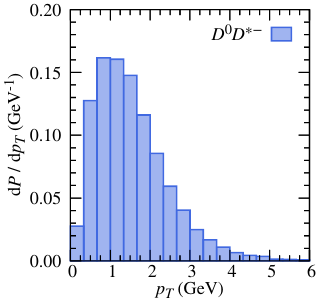}
\includegraphics[scale=0.75]{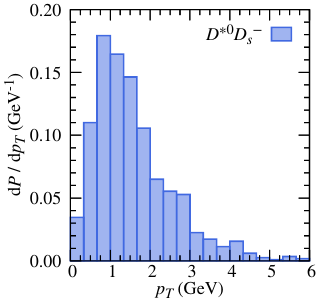}
\caption{Transverse momentum distributions of $J/\psi \pi^-$, $D^0D^{\ast -}$, $J/\psi K^-$ and $D^{\ast 0}D_s^-$ hadron pairs. The distributions are simulated by \textsc{Pythia 8} and normalized such that $\int_0^{\infty}\text{d}p_T\left(\text{d}P/\text{d}p_T\right) =1$.}
\label{fig:pT}
\end{figure}

For $D^{0} D^{\ast -}$ and $D^{\ast 0} D_s^-$ pairs, we select pairs with relative momentum $k<0.4$ GeV in their CM frame, while the $J/\psi \pi^-$ and $J/\psi K^-$ pairs are selected with $k<0.8$ GeV to include the momentum close to the $D^{0} D^{\ast -}$ and $D^{\ast 0} D_s^-$ thresholds. The transverse momentum $p_T$ of each hadron pair in the lab frame is recorded, and the resulting distributions are displayed in Fig.~\ref{fig:pT}. Since the $D^{\ast 0} D^-$ and $D^0 D_s^{\ast -}$ distributions are nearly identical to those of $D^{0} D^{\ast -}$ and $D^{\ast 0} D_s^{-}$, respectively, they are omitted from the figure. From these distributions, we estimate the average transverse mass $\langle m_T\rangle$, with $m_T$ defined as
\begin{align}
m_T=\sqrt{p_T^2+m_{\text{av.}}^2},
\end{align}
where $m_{\text{av.}}$ is the average mass of the hadron pair.

\begin{table}[t]
\caption{\label{Tab:parameter} Average transverse masses and the corresponding core and effective source radii for the channels studied in this work in $pp$ collisions at $\sqrt{s}=13$ TeV. The uncertainties in the source radii are propagated from the parameter errors of the polynomial fit employed to parameterize the source function according to the $\langle m_T\rangle$ scaling of the $pp$ CF reported in Ref.~\cite{ALICE:2020ibs}.}
\renewcommand{\arraystretch}{1.2}
\begin{ruledtabular}
\begin{tabular}{lcccc}
      & $J/\psi \pi^-$ & $D^0 D^{\ast -}$  & $J/\psi K^- $ & $D^{\ast 0} D_s^{-} $ \\[3pt]
\hline
$\left\langle m_T\right\rangle$ [GeV]   & 2.229  & 2.486 & 2.283 & 2.548
       \\[3pt]
$R_{\text{core}}$ [fm]  & $0.79(6)$  & $0.77(7)$ & $0.79(6)$ & $0.76(7)$
       \\[3pt]
$R_{\text{eff}}$ [fm]   & $0.87(6)$  & $0.84(7)$ & $0.86(6)$ & $0.84(7)$
       \\[3pt]
\end{tabular}
\end{ruledtabular}
\end{table}

The resulting transverse masses are collected in Table~\ref{Tab:parameter}, together with the corresponding source radii. The core radius $R_{\text{core}}$ follows from the $pp$ $\langle m_T\rangle$ polynomial fit~\cite{ALICE:2020ibs}, and is enlarged by $\sim 10\%$ to account for feed-down from primordial resonances, giving the effective radius $R_{\text{eff}}$. The resulting radii lie below the typical $1$~fm interaction range, and the $D^{\ast 0}D^-$ and $D^{0}D_s^{\ast -}$ radii coincide with those of $D^{0}D^{\ast -}$ and $D^{\ast 0}D_s^{-}$, respectively. In particular, we find $R_\text{eff}=0.84(7)\,\text{fm}$ for these channels, which we use in order to provide realistic predictions for the CFs, while representative values of $2$ and $5$~fm are adopted for $pA$ and $AA$ collisions, respectively.

The relative production weights $w_j$ are the remaining non-interaction input. Inclusive $pp$ data at $\sqrt{s}=13$~TeV indicate that $J/\psi$ production is much smaller than open-charm production at mid-rapidity~\cite{ALICE:2023sgl}. For the low-relative-momentum pairs relevant here, forming $J/\psi\pi$ or $J/\psi K$ additionally requires the $c\bar c$ pair to form a compact charmonium state, whereas no analogous constraint applies to open-charm meson pairs. Thus, we set the $J/\psi\pi$ ($J/\psi K$) weight to zero, $w_1=0$. Charge-conjugation and light-flavor isospin/SU(3) symmetries imply identical production rates for the $D^0D^{\ast -}$ and $D^{\ast 0}D^-$ pairs, as well as for the $D^{\ast 0}D_s^-$ and $D^0D_s^{\ast -}$ pairs. We therefore take $w_2=w_3=1$. As a check, setting $w_1=1$ produces negligible changes in the central CFs. This reflects the strong suppression of the $J/\psi\pi^-\to D^0D^{\ast -}$ and $J/\psi K^-\to D^{0}D_s^-$ transitions with the LECs of Ref.~\cite{Du:2022jjv} (see the squared $T$-matrix elements in the Supplemental Material~\cite{supp}, Fig.~\ref{fig:Tsq}), together with the additional suppression from the source function, since the momenta at the $D^0D^{-}$ and $D^{\ast 0}D_s^-$ thresholds exceed $600\,\text{MeV}$.

\appsection{Dependence on $C_{1X}$ of the scattering lengths, $W_{c1}$ position, and the CF at threshold}\label{app:scattering-lengths}

\begin{figure}[t]\centering
\includegraphics[scale=1.05]{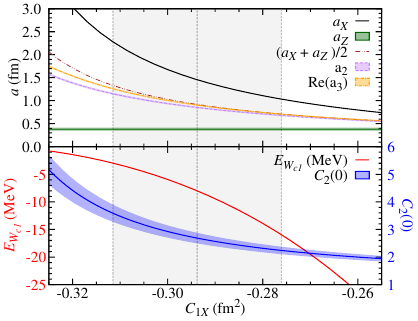}
\caption{Dependence of the non-strange scattering lengths (upper panel), and of the $W_{c1}$ pole position and the threshold CF (lower panel), on the $C$-even LEC $C_{1X}$ for the resonance scenario of the $Z_c(3900)$. Upper panel: eigen-channel scattering lengths $a_X$ (black solid), $a_Z$ (green dashed), and their average $(a_X+a_Z)/2$ (brown double-dot-dashed), together with the physical scattering lengths $a_2$ (violet dashed) and $\mathrm{Re}(a_3)$ (orange dash-dotted). Lower panel: the $W_{c1}$ pole position $E_{W_{c1}}$, measured relative to the $D^0D^{\ast-}$ threshold (red solid, left axis), and the threshold CF $C_{D^0D^{\ast-}}(0)$ (blue solid, right axis). The bands represent the uncertainties associated with the source radius, the remaining LECs, and the UV cutoff $\Lambda$, excluding that of $C_{1X}$. Since $a_X$ and $E_{W_{c1}}$ are determined solely by $C_{1X}$, no uncertainty bands are shown for them. The shaded rectangle indicates the $1\sigma$ interval of $C_{1X}$, $C_{1X}=-0.294(18)\,\mathrm{fm}^2$.\label{fig:run-c1x}}
\end{figure}
\begin{table}[t]
\caption{\label{tab:scattering-lengths}%
$S$-wave scattering lengths (in fm), defined in Eq.~\eqref{eq:SL-definitions}, for the central value $C_{1X}=-0.294\,\mathrm{fm}^2$. Results are shown for the non-strange and strange systems in both the virtual-state and resonance scenarios for the $Z_{c(s)}$.%
}
\renewcommand{\arraystretch}{1.2}
\begin{ruledtabular}
\begin{tabular}{ccccc}
 & \multicolumn{2}{c}{$D^0 D^{\ast-}-D^{\ast0} D^{-}$} & \multicolumn{2}{c}{$D^0 D^{\ast-}_s-D^{\ast0} D^{-}_s$} \\
 & Virtual state & Resonance & Virtual state & Resonance \\
\hline
$a_2$ & $0.902(34)$ & $0.853(19)$ & $0.948(38)$ & $0.892(21)$\\
$\operatorname{Re}a_3$ & $0.940(30)$ & $0.905(17)$ & $0.966(31)$ & $0.933(18)$\\
$\operatorname{Im}a_3$ & $0.062(7)$ & $0.070(4)$ & $0.079(9)$ & $0.090(6)$\\
$a_Z$ & $0.454(56)$ & $0.381(32)$ & $0.469(59)$ & $0.395(33)$ \\
$a_X$ & \multicolumn{2}{c}{$1.457$} & \multicolumn{2}{c}{$1.550$} \\
\end{tabular}
\end{ruledtabular}
%\end{tabularx}
\end{table}

The LEC $C_{1X}=-0.294(18)\,\mathrm{fm}^2$ is fixed by reproducing the predicted $W_{c1}$ pole, located $8^{+8}_{-5}\,\text{MeV}$ below the $D^0D^{\ast-}$ threshold~\cite{Zhang:2024fxy,Ji:2025hjw}, which implies a pole $7^{+7}_{-5}\,\text{MeV}$ below the $D^{\ast0}D_s^-$ threshold. Owing to the shallow virtual nature of the $W_{c1}$, threshold observables exhibit a strong, nonlinear dependence on $C_{1X}$. Rather than propagating its uncertainty into the CF predictions, we explicitly investigate the dependence of the CF at threshold and related observables on $C_{1X}$.

We define the physical scattering lengths as:
\begin{equation}\label{eq:SL-definitions}
a_{i} \equiv -\frac{T_{ii}(E_{\text{th},i})}{8\pi \,E_{\text{th},i}}\,,
\end{equation}
with $i=2,3$ referring to $D^0 D^{\ast-}$ and $D^{\ast 0}D^{-}$, respectively, and $E_{\text{th},i}$ the threshold CM energy of channel $i$. For the $C$-even and $C$-odd scattering lengths, denoted as $a_{X}$ and $a_{Z}$, we use the same convention as Eq.\,\eqref{eq:SL-definitions}, and compute the amplitude in a single-channel approximation, neglecting mass splittings, and with a potential given only by $C_{1X}$ and $C_{1Z}$, respectively. The imaginary parts of $a_X$ and $a_Z$ vanish by construction, while that of $a_2$ is negligible owing to the weak coupling to the $J/\psi\pi$ channel. The same definitions apply to the strange channels, whose corresponding states have $I=1/2$, and we retain the notation $a_{(X,Z)}$ for simplicity.

For the central value $C_{1X}=-0.294\,\mathrm{fm}^2$, the corresponding scattering lengths are listed in Table~\ref{tab:scattering-lengths}. Figure~\ref{fig:run-c1x} shows the dependence of the $W_{c1}$ pole position (measured relative to the $D^0D^{\ast-}$ threshold), the scattering lengths $a_2$, $a_3$, $a_{X}$, and $a_{Z}$, and the threshold value $C_{D^0D^{\ast-}}(0)$ on the LEC $C_{1X}$. Neglecting the mass splitting between the two hidden-charm thresholds, one obtains, in analogy with Eq.~\eqref{eq:relation-CFs-C-parity},
\begin{equation}\label{eq:SL-average}
a_2=a_3=\frac{a_{X}+a_{Z}}{2}\,.
\end{equation}
This relation is well satisfied by the results in Table~\ref{tab:scattering-lengths} and Fig.~\ref{fig:run-c1x}, except when the $W_{c1}$ pole lies very close to the $D^0D^{\ast-}$ threshold, where isospin-breaking effects from the threshold splitting become more pronounced. The relation provides a simple illustration of the central result of this work: the physical observables receive comparable contributions from both the $C$-even and $C$-odd interactions, making it essential to treat them simultaneously.

As the interaction becomes more attractive (moving leftwards in $C_{1X}$), the virtual $W_{c1}$ pole approaches the $D^0D^{\ast-}$ threshold, causing $a_X$, $a_{2,3}$, and $C_{D^0D^{\ast-}}(0)$ to increase accordingly. By construction, $a_Z$ is independent of $C_{1X}$ and thus provides a natural reference. This pronounced sensitivity explains why the uncertainty in $C_{1X}$ is not propagated into the CF predictions of Fig.~\ref{fig:CF_predict}. Conversely, a measurement of the threshold CF would tightly constrain $C_{1X}$, providing a direct determination of the $W_{c1}$ pole position.

\onecolumngrid
\clearpage

\setcounter{figure}{0}
\setcounter{section}{0}
\setcounter{equation}{0}
\setcounter{table}{0}
\setcounter{page}{1}
\setcounter{secnumdepth}{1}
\makeatletter 
\renewcommand{\thefigure}{S\@arabic\c@figure}
\renewcommand{\thetable}{S\@arabic\c@table}
\renewcommand{\thesection}{S\@arabic\c@section}
\renewcommand{\theequation}{S\@arabic\c@equation}
\makeatother

\titleformat{\section}{\centering\small\bfseries\MakeUppercase}{\thesection.}{5pt}{}[]
\titlespacing*{\section}
{0pt}{18pt plus 1pt minus 1pt}{9pt plus 1pt}

\begin{onecolumngrid}

\begin{center}
{\uppercase{\bfseries\normalsize Supplemental Material}}
\end{center}

\section[Comparison with the single-channel results of Z.-W.\,Liu et al.]{Comparison with the single-channel results of Z.-W.\,Liu \textit{et al.}}\label{supp-mat:comparisons}

Figure~\ref{fig:CF_single} compares the single-channel CFs of the three scenarios considered in Ref.~\cite{Liu:2024nac} (dark-blue solid lines) with the corresponding coupled-channel results obtained for $C_{1X}=-0.294\,\mathrm{fm}^2$ (green dashed lines) and $C_{1X}=0$ (dash-dotted lines). For a direct comparison, the latter are computed using the potential of Eq.~\eqref{eq:V1Z-energy}, with $C_{1Z}$ and $b$ adjusted to reproduce the pole positions adopted in Ref.~\cite{Liu:2024nac}. The double-dot-dashed lines show the combination $[1+C^{\mathrm{Liu}}(k)]/2$, where $C^{\mathrm{Liu}}$ denotes the single-channel CFs of Ref.~\cite{Liu:2024nac}. This follows from Eq.~\eqref{eq:relation-CFs-C-parity} of the main text: for degenerate thresholds and equal production weights, the physical CF is the average of the two $C$-parity eigenchannel CFs. At $C_{1X}=0$, the $C$-even interaction vanishes, implying $\widetilde{C}_{1X}(k)=1$ and hence $C_{D^0D^{\ast-}}(k)=[1+\widetilde{C}_{1Z}(k)]/2$. Identifying $\widetilde{C}_{1Z}(k)$ with the single-channel result of Ref.~\cite{Liu:2024nac} therefore yields the physical CF corresponding to their interaction, which can be directly compared with our $C_{1X}=0$ results.

\begin{figure}[b]\centering
\includegraphics[scale=0.80]{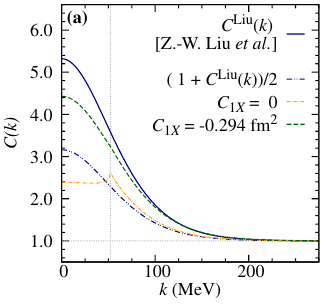}%
\includegraphics[scale=0.80]{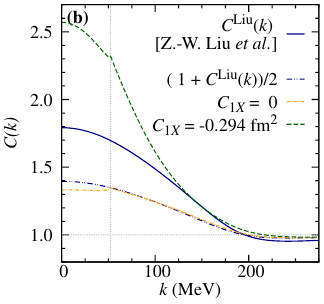}%
\includegraphics[scale=0.80]{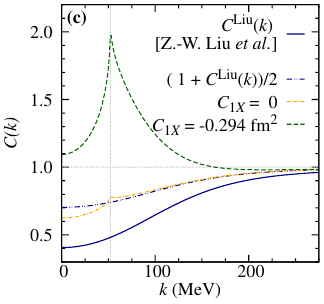}\\
\includegraphics[scale=0.80]{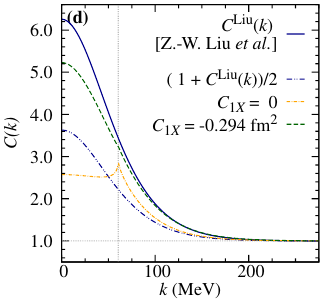}%
\includegraphics[scale=0.80]{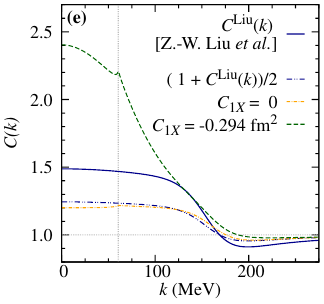}%
\includegraphics[scale=0.80]{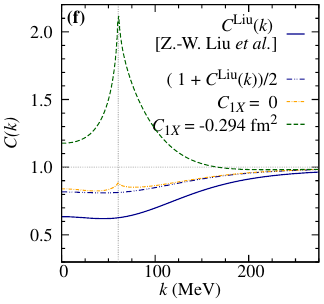}
\caption{CFs for the $D^0 D^{\ast -}$ (upper) and $D^{\ast 0} D_s^-$ (lower) channels as a function of the relative CM momentum $k$. The left, middle, and right columns correspond to the virtual-, resonant-, and bound-state scenarios, respectively. The dark-blue solid lines show the single-channel results (central curves reported in Fig.~2 of Ref.~\cite{Liu:2024nac}), and the dark-blue double-dot-dashed line corresponds to the observable combination described in the text. The orange dash-dotted and green dashed lines denote the coupled-channel results with $C_{1X}=0$ and $C_{1X}=-0.294\,\text{fm}^2$, respectively. The latter scenario accounts for the effects induced by the isovector partner $W_{c1}$ of the $X(3872)$~\cite{Zhang:2024fxy}. The vertical lines indicate the $D^{\ast 0}D^-$ and $D^0 D_s^{\ast -}$ thresholds.}
\label{fig:CF_single}
\end{figure}
The two are indeed in excellent agreement over the entire momentum range, supporting the interpretation advanced in the main text that the results of Ref.~\cite{Liu:2024nac} should be identified with $\widetilde{C}_{1Z}(k)$ rather than with the CFs of the physical charge states to which they are assigned. This agreement relies on two conditions.
\begin{itemize}
\item First, the averaging relation must hold. It is exact only in the degenerate-threshold limit, while the $1.3$ and $1.8$~MeV threshold splittings in the non-strange and strange systems introduce small violations. Accordingly, the coupled-channel results develop cusps at the $D^{\ast0}D^-$ and $D^0D_s^{\ast-}$ thresholds that are absent in the single-channel treatment and cannot be reproduced by construction by the combination $[1+C^{\mathrm{Liu}}]/2$. The line shapes of such near-threshold structures in symmetry-related two-channel systems were classified in Ref.~\cite{Zhang:2024qkg} in terms of the single-channel scattering length and the channel-coupling strength, which govern the pole trajectories and thus the evolution from a threshold cusp to a below-threshold peak. In the present case, the $C$-even LEC $C_{1X}$ controls the coupling between the two charge channels and thereby reshapes these threshold structures.

\item Second, our $C$-odd interaction and that of Ref.~\cite{Liu:2024nac} must yield similar scattering amplitudes. This requirement is fulfilled because our $C_{1X}=0$ (orange, single-dot-dashed) curves are adjusted to reproduce the same $C$-odd pole positions in the different scenarios. However, reproducing the same pole position does not uniquely determine the amplitude away from the pole, so small residual differences are unavoidable.
\end{itemize}

As anticipated, Fig.~\ref{fig:CF_single} shows that the differences between $[1+C^{\mathrm{Liu}}(k)]/2$ and our $C(k)$ obtained with $C_{1X}=0$ are indeed small. They nevertheless exhibit a clear systematic pattern across the three scenarios: they are smallest in the resonance case and largest in the virtual-state one. This ordering reflects the distance of the pole from threshold. Using the masses of Ref.~\cite{Liu:2024nac}, the virtual state lies about $4$~MeV below the $D^0D^{\ast-}$ threshold, the bound state about $8$~MeV below, and the resonance about $12$~MeV above it. The closer the pole is to threshold, the larger the threshold enhancement of the CF, which scales quadratically with the scattering length. Consequently, any residual difference between two amplitudes sharing the same pole is amplified. This explains why the virtual-state scenario exhibits the largest discrepancy, whereas the resonance scenario shows almost perfect agreement.

By contrast, these results differ markedly from the curves obtained with $C_{1X}=-0.294\,\mathrm{fm}^2$, which incorporate the effects of the predicted $W_{c1}$ state. This further reinforces the main conclusion of our work: the $C$-even interaction induces sizable modifications of the $D^0D^{\ast-}{(s)}$ and $D^{\ast0}D^{-}{(s)}$ CFs and must therefore be taken into account.

\section{Individual-channel contributions to CFs}\label{supp-mat:contributions}

We define the contribution of channel $j$ to the CF of channel $i$ as
\begin{align}
{C}_{ji}(k)=
4\pi\int_0^{+\infty}dr\, r^2S_{j}(r) \left(w_j|\tilde{\psi}_{ji}(k,r)|^2-j_0^2(kr)\delta_{ij}\right).
\end{align}
The individual contributions $C_{ji}$ to the $D^0D^{\ast-}$ and $D^{\ast0}D_s^-$ CFs, computed with the source radii of Table~\ref{Tab:parameter}, are shown in the upper panels of Fig.~\ref{fig:Cij} for the equal-weight case, $w_1=w_2=w_3=1$. In the calculations presented in the main text, however, $w_1=0$, so that $C_{1i}=0$; restoring $w_1=1$ changes the total CFs by less than $3\times10^{-5}$ over the momentum range considered. The diagonal contributions, $D^0D^{\ast-}\to D^0D^{\ast-}$ and $D^{\ast0}D_s^-\to D^{\ast0}D_s^-$ ($C_{22}$), dominate because of the large magnitude of $T_{22}$. As shown in the upper panel of Fig.~\ref{fig:Tsq}, $|T_{22}|^2$ is substantially larger than $|T_{21}|^2$ and $|T_{23}|^2$ in both the $J/\psi\pi^-$--$D^0D^{\ast-}$--$D^{\ast0}D^-$ and $J/\psi K^-$--$D^{\ast0}D_s^-$--$D^0D_s^{\ast-}$ coupled-channel systems. Although $C_{32}$ is smaller than $C_{22}$, it still gives a visible contribution.

When $C_{1X}=0$, $|T_{22}|^2$ becomes identical\footnote{One finds $T^{-1}_{22}=-T^{-1}_{23}=2/C_{1Z}^{\prime}-(G_2+G_3)$, where $T^{-1}_{ij}\equiv 1/T_{ij}$ denotes the reciprocal of the matrix element: at $C_{1X}=0$ the $2\times2$ block is rank one and is not invertible.} to $|T_{23}|^2$ (see lower panels of Fig.~\ref{fig:Tsq}) and differs significantly from the results shown in the upper panels, obtained using $C_{1X}=-0.294\,\mathrm{fm}^2$, determined as discussed in the main text. Note that the $y$-axis scales in the upper and lower panels differ by a factor of 10. Nevertheless, the interference between the spherical Bessel function and the $T_{22}$ term in the wave function of Eq.~\eqref{eq:wave_fuc} leads to $C_{22}>C_{23}$ in the low-momentum region, as shown in the lower panel of Fig.~\ref{fig:Cij}.

Notably, the interaction in the $J/\psi\pi \to D\olsi D{}^\ast /D^\ast \olsi D$ channel is much weaker than that in the $D\olsi D{}^\ast /D^\ast \olsi D\to D\olsi D{}^\ast /D^\ast \olsi D$ channel~\cite{Du:2022jjv}. This conclusion differs from the HAL QCD results~\cite{Ikeda:2017mee}, possibly because the latter was obtained with unphysically heavy pion masses, $m_{\pi}=410-700$~MeV. This hierarchy of interactions is supported by the LECs determined from a systematic reanalysis of the recently BESIII data~\cite{Chen:2023def,Chen:2026fnz}. Future lattice QCD simulations at or near the physical pion mass will provide a decisive test of this picture.

\begin{figure}[t]
\centering
\includegraphics[scale=0.75]{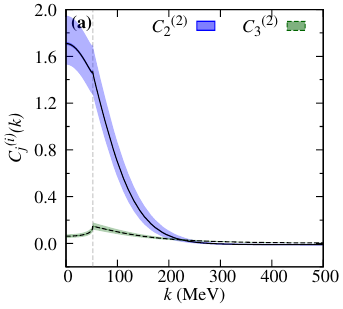}%
\includegraphics[scale=0.75]{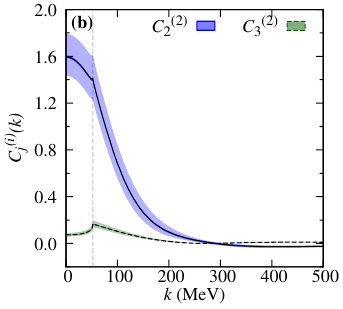}%
\includegraphics[scale=0.75]{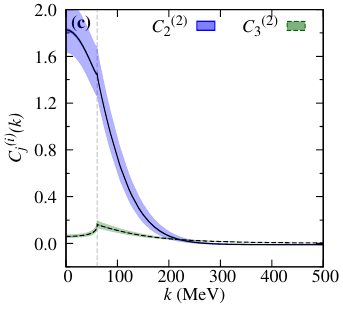}%
\includegraphics[scale=0.75]{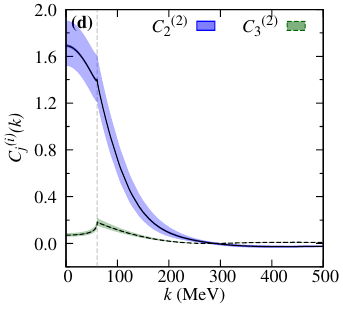}\\
\includegraphics[scale=0.75]{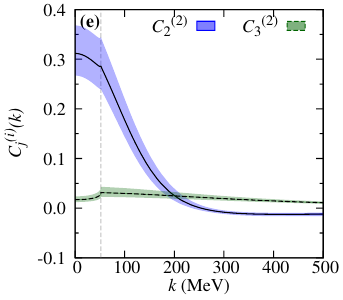}%
\includegraphics[scale=0.75]{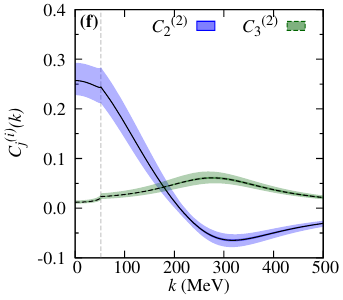}%
\includegraphics[scale=0.75]{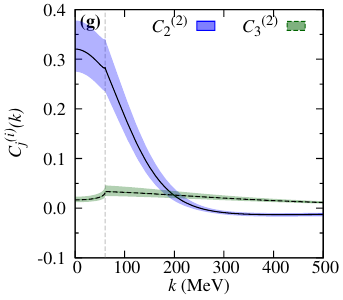}%
\includegraphics[scale=0.75]{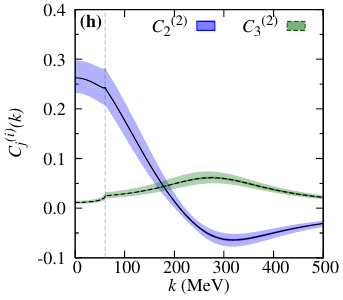}
\caption{Contributions of individual channels to $D^{0} D^{\ast -}$ and $D^{\ast 0}D_s^-$ CFs in the $J/\psi\pi^-$--$D^0 D{}^{\ast -}$--$D{}^{\ast 0}D^-$ and $J/\psi K^-$--$D{}^{\ast 0} D_s^-$--$D{}^{0} D_s^{\ast -}$ coupled-channel systems. The first (last) two columns correspond to the $D^{0} D^{\ast -}$ ($D^{\ast 0}D_s^-$) CFs. Within each system, the left and right panels represent the virtual-state and resonant-state scenarios, respectively. The upper and lower panels are obtained with $C_{1X}=-0.294\,\text{fm}^2$ and $C_{1X}=0$, respectively. The vertical lines indicate the $D^{\ast 0}D^-$ and $D^0 D_s^{\ast -}$ thresholds.}
\label{fig:Cij}
\end{figure}

\begin{figure}[t]
\centering
\includegraphics[scale=0.75]{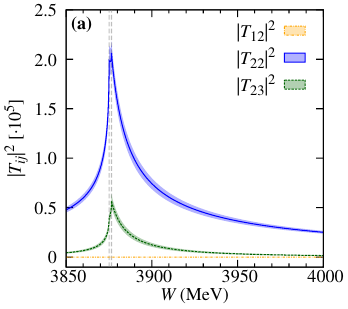}%
\includegraphics[scale=0.75]{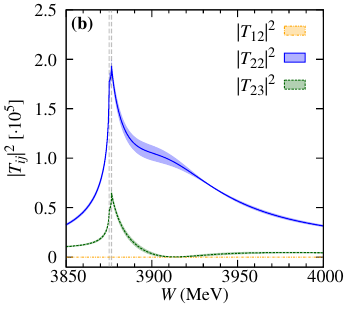}%
\includegraphics[scale=0.75]{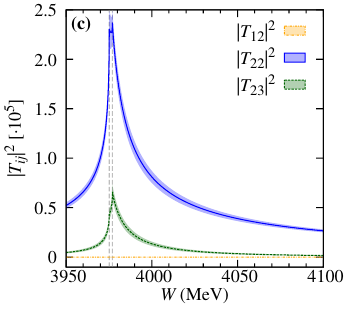}%
\includegraphics[scale=0.75]{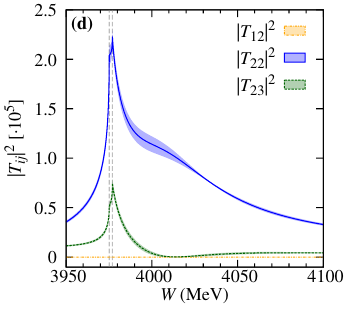}\\
\includegraphics[scale=0.75]{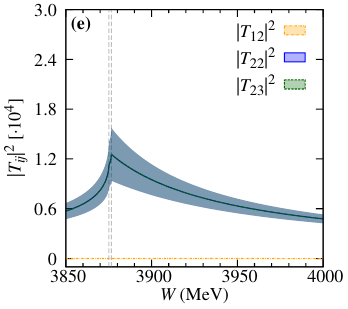}%
\includegraphics[scale=0.75]{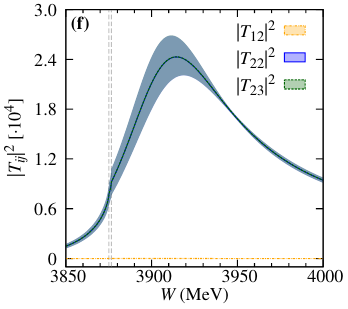}%
\includegraphics[scale=0.75]{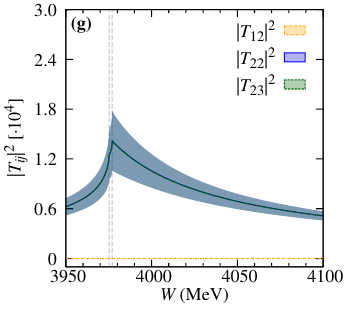}%
\includegraphics[scale=0.75]{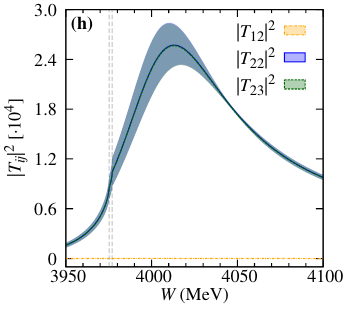}
\caption{Line shapes of $J/\psi\pi^-$--$D^0 D{}^{\ast -}$--$D{}^{\ast 0}D^-$ and $J/\psi K^-$--$D{}^{\ast 0} D_s^-$--$D{}^{0} D_s^{\ast -}$ coupled-channel systems. The labels (a)--(h) denote the same cases as in Fig.~\ref{fig:Cij}. The vertical lines indicate the two closely spaced thresholds, $D^0 D{}^{\ast -}$--$D{}^{\ast 0}D^-$ and $D{}^{\ast 0} D_s^-$--$D{}^{0} D_s^{\ast -}$, depending on the case.}
\label{fig:Tsq}
\end{figure}

\begin{figure}\centering
\includegraphics[scale=0.75]{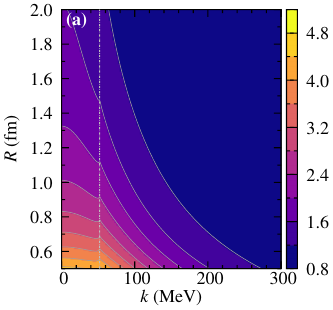}%
\includegraphics[scale=0.75]{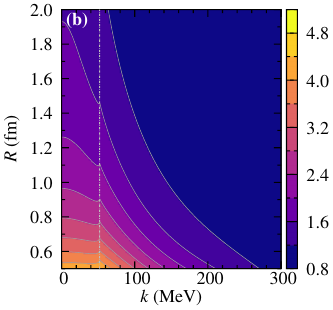}%
\includegraphics[scale=0.75]{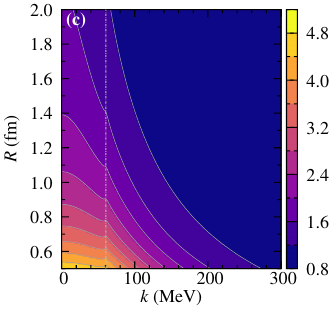}%
\includegraphics[scale=0.75]{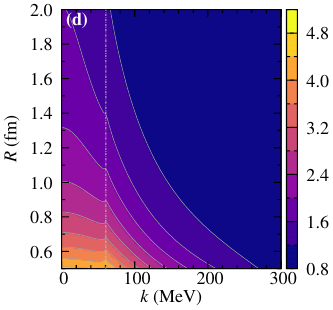}
\caption{CFs for the $D^0D^{\ast -}$ and $D^{\ast 0}D_s^{-}$ pairs as a function of the relative momentum $k$ and the $pp$ source radius. The labels (a)--(d) denote the same cases as in Fig.~\ref{fig:Cij}. The white vertical lines indicate the positions of the $D^{\ast 0}D^{-}$ and $D^{0}D_s^{\ast -}$ thresholds.}
\label{fig:C_R}
\end{figure}

Finally, to investigate the impact of the source radius on the CFs, we fix the effective $pp$ source radii for the $J/\psi\pi^-$ and $J/\psi K^-$ channels to the values given in Table~\ref{Tab:parameter}, due to their negligible contributions, and vary the source radii of the remaining channels from $0.5$~fm to $2.0$~fm. The relative momentum is taken in the range $k\in[0,0.5]$~GeV. As shown in Fig.~\ref{fig:C_R}, the CFs at small relative momentum ($k<0.2$~GeV) depend strongly on the source radius $R$. In particular, for $k<0.1$~GeV, the $D^0D^{\ast -}$ and $D^{\ast 0}D_s^{-}$ CFs exhibit a pronounced sensitivity to $R$ when $R<1.0$~fm, in both the virtual-state and resonance scenarios. This behavior highlights that an accurate determination of the source radius under the given Gaussian profile assumption is essential in femtoscopic studies to reliably extract low-energy scattering information~\cite{Epelbaum:2025aan,Molina:2025lzw}.

\end{onecolumngrid} 

\end{document}